\documentclass[twocolumn]{aastex63} %twocolumn

\usepackage[version=4]{mhchem}
\usepackage{graphicx}

\usepackage{longtable}

\newcommand{\jwst}{{\it JWST}}

\newcommand{\lacc}{$L_{acc}$ }
\newcommand{\mic}{$\mu$m }

\shorttitle{Molecular Gas Variability in DQ Tau}
\shortauthors{Hyden et al.}

\begin{document}

\title{Coordinated Space and Ground Monitoring of Accretion Bursts in a Protoplanetary Disk: \\ MIRI and iSHELL observe the long-term modulation of gas within a small inner disk cavity}

\correspondingauthor{}
\email{ehd28@txstate.edu, banzatti@txstate.edu}

\author[0009-0008-2382-0614]{Jacob Hyden}
\affiliation{Department of Physics, Texas State University, 749 N Comanche Street, San Marcos, TX 78666, USA}

\author[0000-0003-4335-0900]{Andrea Banzatti}
\affiliation{Department of Physics, Texas State University, 749 N Comanche Street, San Marcos, TX 78666, USA}

\author[0000-0003-2053-0749]{Benjamin M.\ Tofflemire}
\affiliation{SETI Institute, Mountain View, CA 94043, USA/NASA Ames Research Center, Moffett Field, CA 90345 USA}
\affiliation{Department of Astronomy, The University of Texas at Austin, Austin, TX 78712, USA}

\author[0000-0002-3291-6887]{Sebastiaan Krijt}
\author[0000-0001-8240-978X]{Till Kaeufer}
\affiliation{Department of Physics and Astronomy, University of Exeter, Stocker Road, Exeter EX4 4QL, UK}

\author[0000-0001-7552-1562]{Klaus M.\ Pontoppidan}
\affiliation{Jet Propulsion Laboratory, California Institute of Technology, 4800 Oak Grove Drive, Pasadena, CA 91109, USA}

\author[0000-0003-3562-262X]{Carlo F.\ Manara}
\affiliation{European Southern Observatory, Karl-Schwarzschild-Strasse 2, 85748 Garching bei München, Germany}

\author[0000-0003-1817-6576]{Christian Rab}
\affiliation{University Observatory, Faculty of Physics, Ludwig-Maximilians-Universität München, Scheinerstr. 1, D-81679 Munich}
\affiliation{Max-Planck-Institut für extraterrestrische Physik, Giessenbachstrasse 1, D-85748 Garching, Germany}

\author[0009-0004-8022-8597]{John M. McCaskill}
\affiliation{Department of Physics, Texas State University, 749 N Comanche Street, San Marcos, TX 78666, USA}

\author[0000-0002-6777-6386]{Guillaume Bourdarot}
\affiliation{Max-Planck-Institut für extraterrestrische Physik, Giessenbachstrasse 1, D-85748 Garching, Germany}

\author[0000-0002-3741-9353]{Emma T.\ Whelan}
\affiliation{Department of Physics, Maynooth University, Maynooth, Co. Kildare, Ireland}

\author[0000-0002-5758-150X]{Joan Najita}
\affiliation{NSF’s NOIRLab, 950 N. Cherry Avenue, Tucson, AZ 85719, USA}

\author[0000-0001-8790-9011]{Adrien Houge}
\affiliation{Center for Star and Planet Formation, Globe Institute, {\O}ster Voldgade 5, 1350 Copenhagen, Denmark}

\author[0000-0001-9344-0096]{Adwin Boogert}
\affiliation{Institute for Astronomy, University of Hawaii, 2680 Woodlawn Drive, Honolulu, HI, 96822 USA}

\author[0000-0001-6660-936X]{Catherine Dougados}
\affiliation{Université Grenoble Alpes, CNRS, IPAG, 38000 Grenoble, France}

\author[0000-0002-7695-7605]{Myriam Benisty}
\affiliation{Max-Planck Institute for Astronomy (MPIA), Königstuhl 17, 69117 Heidelberg, Germany}

\author[0000-0002-5943-1222]{James Muzerolle Page}
\affiliation{Space Telescope Science Institute, 3700 San Martin Dr, Baltimore, MD 21218, USA}

\begin{abstract}

DQ~Tau is a TTauri eccentric binary with pulsed accretion bursts regulated by its $\sim 16$-day orbital period. We have monitored this system with seven instruments from ground and space over 40\,days. We present results on molecular gas variability from nine JWST-MIRI and nine IRTF-iSHELL epochs over three sequential accretion bursts with accretion luminosity pulses increasing by factors of $\sim 9$--80. A fundamental new measurement from this analysis is the line broadening of CO (from iSHELL) and \ce{H2O} (from MIRI) lines, which we use to estimate Keplerian radii and their variability in parallel to the line luminosity. This unprecedented dataset clearly shows that the molecular emission does not react like HI and the 5-$\mu$m continuum with each accretion burst: the line luminosity and width of all hot molecular gas tracers decrease over $\sim30$--35~days, consistent with an emitting region decreasing in temperature and mass as it recedes to larger radii within the $\sim 0.2$--0.4\,au inner cavity perturbed by the binary. The infrared molecular spectrum of DQ~Tau is affected more by disk dynamics then by chemical changes triggered by variable accretion, supporting model expectations for chemical relaxation timescales that are much longer than the duration of these accretion bursts ($\gg 10$~days). The variability observed in all tracers in this study is consistent with a scenario where gas mass builds up near the disk edge to modulate the burst peak intensity and gas density in the inner cavity every $\sim4$--5 orbital periods. The origin of this modulation is still unclear.

\end{abstract}

\section{Introduction} \label{sec: intro}
The chemical composition of gas and dust in the inner regions of protoplanetary disks has been proposed to evolve during planet formation through multiple processes. With inner disks being largely shaped by radial and vertical gradients in temperature and density, a number of factors may impact disk chemistry beyond what is expected in a static, smooth disk geometry: disk structures like radial or azimuthal over-densities and gaps \citep{andrews20,bae23}, gas dispersal through disk winds \citep{alexander14,pascucci23}, planet formation \citep{Drazkowska23}, irradiation conditions that slowly evolve over Myr timescales \citep[][]{Manara23}, and even the sudden irradiation changes produced by episodic accretion outbursts \citep{Audard14,fischer23}. While during the T~Tauri phase the average accretion variability is about a factor of 2--3 (0.3--0.5 dex) over timescales as short as days and as long as years \citep{costigan14,venuti14}, episodic ``bursters" have been observed to change by factors of 10--100 over timescales from days to weeks and in exceptional cases up to several months \citep[the EX~Lup outbursts and V1647~Ori outbursts,][]{fischer23}. These stronger events, while shorter and still moderate in comparison to the FU~Ori-type outbursts happening at earlier stages, have been observed to significantly change the dust and gas emission from inner disks during the planet formation phases at $\sim 1$~Myr \citep{abraham09,banz12,2025smith}. In fact, the increased UV and X-ray irradiation during outbursts is expected to alter the inner disk chemistry, with potential longer-term impact on planet formation. These effects include: UV photo-dissociation of molecules, increased molecular formation in a warmer disk region, ice sublimation during higher accretion events \citep{cleeves17,molyarova18,waggoner19,Woitke2024,houge2023,ros2024,2025smith}. All these effects can have an observable impact by increasing or decreasing the emission from a given molecule, depending on the case.

While episodic accretion bursts are thought to be common in young stars as they assemble their mass, their low frequency and unpredictability make them hard to use for studies of inner disk chemistry during planet-formation in the Class II phase of solar-mass stars, which have so far been rare and limited to a few epochs. Pulsed accretion in disks around binary systems offer an alternative predictable experiment to monitor inner disk chemistry as a function of variable irradiation conditions. However, these systems also have complex inner disk dynamics where accretion onto the stars is regulated by the interactions between the inner disk and the binary orbit, producing asymmetric and highly variable structures at the inner edge \citep{Artymowicz96,2016MunozR,2012shi,2015dunhill,2002guntherkeys,val-borro2011,2011sytov}. The interactions between physics, chemistry, and variable irradiation conditions in these inner disks are still mostly unexplored.

With the launch of the James Webb Space Telescope (JWST), the community has regained spectroscopic access to study the inner disk within a few astronomical units from the star with higher sensitivity and resolving power over a similar spectral range previously provided by the Infrared Spectrograph (IRS) on the Spitzer Space Telescope. The Mid-Infrared Instrument (MIRI) aboard JWST has now reopened the opportunity for time-domain studies of mid-infrared molecular gas variability in connection to accretion variability \citep{2025smith,kospal25}. In this paper, we report on the analysis of inner disk molecular gas variability observed in the binary pulsed-accretor, DQ~Tau, as part of a series of papers dedicated to a multi-observatory monitoring campaign of this system \citep{Tofflemire25,2025Hala}.

\begin{deluxetable}{l c c}
\tablecaption{\label{tab: DQTau props} Properties of DQ~Tau system from the literature.}
\tablehead{Parameter & Value & Ref.}
\tablecolumns{3}
\startdata
Distance (pc) & 196 & 1 \\
Total stellar mass ($M_{\odot}$) & $1.21 \pm 0.26$ & 2 \\
Total stellar luminosity ($L_{\odot}$) & $0.64 \pm 0.25$ & 8 \\
Eccentricity $e$ & $0.546 \pm 0.008$ & 3 \\
Orbital Period (days)& $15.71 \pm 0.07$ & 3\\
Stellar RV (km/s) & $20.33 \pm 0.07$ & 3 \\
Millimeter disk inclination (deg) & $19.6 \pm 0.1$ & 4 \\
Outer dust disk radius 90\% (au) & 37 & 5 \\
Binary separation at periastron (au) & 0.063 & 3 \\
Binary separation at apastron (au) & 0.22 & 3 \\
Semi-major axis $a$ (au) & 0.14--0.145 & 3,6 \\
Theoretical inner disk edge $\sim 3a$ (au) & $\sim$~0.42--0.44 &  \\
\enddata
\tablecomments{References: 1) \cite{gaia_mission,gaiaDR3}, 2) \cite{Czekala16}, 3) \cite{2025Hala}, using an orbital inclination of 158 deg, 4) \cite{Yamaguchi24}, 5) \cite{long19}, updated to a distance of 196 pc, 6) Bourdarot et al. in prep, 8) \cite{Muzerolle19}.}
\end{deluxetable}

\begin{deluxetable*}{l c c c c c c c c c c c c c c}
\tabletypesize{\footnotesize}
\tablewidth{0pt}
\tablecaption{Summary of observations obtained in this coordinated monitoring campaign of DQ~Tau in 2024.}
\label{tab: obs_campaign}
\tablehead{ Instrument & Observatory & \# epochs & Dates & Program & PIs & Publications } 
\tablecolumns{7}
\startdata
MIRI & JWST & 9 & Sep 11 -- Oct 19 & GO 4727 & Banzatti, Tofflemire, Rab & 1, 2 \\
iSHELL & IRTF & 9 & Sep 6 -- Oct 17 & 2024B065 & Tofflemire, Banzatti & 2 \\
Sinistro & LCO & 340 & Aug 13 -- Jan 22 & UTX2024B-001 & Tofflemire & 1, 2, 3 \\
X-shooter & VLT & 27 & Sep 5 -- Nov 21 & 114.2799.001 & Manara, Tofflemire & 1, 3 \\
UVES & VLT & 22 & Sep 2 -- Jan 8 & 114.27MG.001 & Whelan & 3, in prep. \\
MUSE & VLT & 2 & Oct 13, Nov 11 & 114.27MG.001 & Whelan & in prep. \\
GRAVITY & VLTI & 8 & Sep 8, Dec 22 -- Jan 7  & 114.27PS.001 & Bourdarot, Benisty & in prep. \\
\enddata
\tablecomments{1: \citet{Tofflemire25}, 2: this work, 3: \citet{2025Hala}. Dates extending to January were in 2025.}
\end{deluxetable*} 

DQ~Tau is a spectroscopic equal-mass binary system in Taurus at a distance of 196~pc (see Table \ref{tab: DQTau props}). This system has long been studied as a prototype of ``pulsed accretion" triggered by the interaction between the stellar orbits and the inner disk \citep{Artymowicz96,Mathieu97}, also observed in UZ~Tau~E and TWA~3A \citep{Jensen2007,2017TofflemireTWA, tofflemire19}. The two stars in DQ~Tau have roughly equal mass of $\sim0.6$~$M_{\odot}$ for a total dynamical mass of $1.21 \pm 0.26 M_{\odot}$, with a binary period of $\sim15.7$ days and eccentricity of $\sim 0.55$ \citep{Mathieu97,Czekala16,2025Hala}. The circumbinary disk is relatively compact, with an outer size of 37~au (including 90\% of the millimeter flux) and possibly a narrow unresolved substructure at $\sim 15$--16~au \citep[][Long et al. in prep.]{long19,zhang23, Yamaguchi24}. The disk inclination from ALMA data in 2016 is $19.6 \pm 0.1$~deg \citep{Yamaguchi24}, confirmed by new higher-resolution ALMA data (F. Long, priv. comm.). The binary orbit has recently been updated in \citet{2025Hala} as part of the coordinated observing campaign described in this work.
It is expected by dynamical models that the resonant gravitational torques from the binary should produce a low-density inner disk region out to $\sim 2$--4 times the semi-major axis $a$ \citep{Artymowicz96,ragusa25, miranda2017}. This corresponds to $\sim 0.3$--0.6~au using semi-major axis estimates reported in Table \ref{tab: DQTau props}, where we adopt the intermediate value of $3a$. In this work, we refer to this inner region as the inner disk ``cavity" produced by the binary dynamics, which we remark it is not expected to be fully devoid of gas, azimuthally symmetric, nor stable in time \citep[e.g.][]{2016MunozR}. Near-infrared interferometry detected dust emission from inside this inner cavity edge \citep{Boden09}, and previous spectrally-resolved $M$-band CO lines also showed large velocities of 70--80~km/s (FWHM), indicating CO emission extending well within this cavity down to $\sim0.1$~au \citep{Carr01}.

The accretion rate and luminosity in DQ~Tau have been recorded to typically vary within one order of magnitude (in extreme cases up to almost two orders of magnitude) with pulses at almost every periastron passage \citep[][and Appendix \ref{app: accretion_history}]{Tofflemire17,Tofflemire25,kospal18,Fiorellino22}. The regular pulses enable the coordination of observations at specific accretion phases to study the impact of accretion variability on the physical and chemical structure of the inner disk. 

In this paper, we report results from the monitoring of molecular gas over three accretion bursts observed in 2024 in DQ~Tau to study their connections with and impact on inner disk physics and chemistry. In Section \ref{sec: results} we analyze the molecular emission as a function of time and variation in \lacc, by including both the observed line luminosity, the radial emitting region as estimated from line widths, and the excitation conditions estimated from multi-component slab fits to the water spectra. We also put the variability measured in DQ~Tau in the broader context of accretion luminosity relations found in a larger sample of disks around single stars. In Section \ref{sec: discussion}, we discuss a potential interpretation that could reconcile all the available observables included in this work, by building on previous results and models. We also discuss the implications of this work for inner disk chemistry and the effects of episodic accretion in general.

\section{Observations} \label{sec: data}
\subsection{Setting up a rare multi-observatory campaign}
\begin{figure*}
\centering
\includegraphics[width=\textwidth]{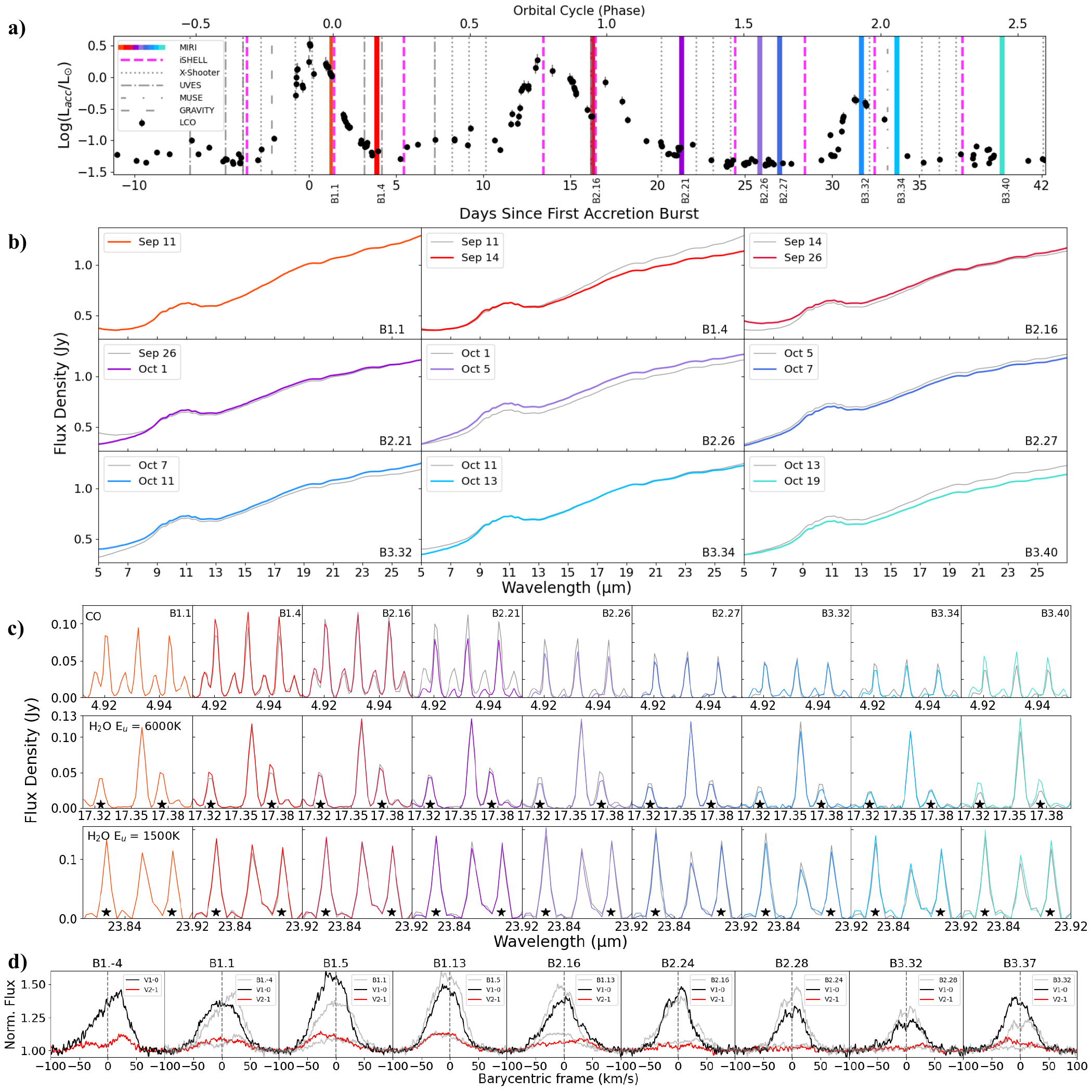}
\caption{ a) LCO light curve observations covering the three accretion bursts in this campaign \citep{Tofflemire25}, with epochs from all instruments marked for reference (see Table \ref{tab: obs_campaign}). b) JWST MIRI continuum as a function of time. c) Variability observed with MIRI in spectral lines from CO and \ce{H2O}, with different energy levels as labeled. d) CO line profiles as observed with iSHELL ($v=1-0$, using line stacks from J = 25--31). All epochs are labeled as ``Bn.days", where ``n" is the burst number (1 to 3) and ``days" is counted from the peak of the first burst (e.g., B2.21 means the second burst has happened and this epoch is 21 days after the first burst). In each plot, the previous epoch of data is shown in light grey for comparison. In the case of water lines, the labeled upper energy level emission are marked by the black stars for identification in each panel. }
\label{fig: main_display}
\end{figure*}

This work is part of a larger multi-instrument campaign that was coordinated across proposals submitted to multiple observatories on the ground and in space (Table \ref{tab: obs_campaign}). In addition to MIRI on JWST (Section \ref{sec: JWST_obs}) and iSHELL on IRTF (Section \ref{sec: iSHELL_obs}), DQ~Tau was observed in multiple epochs between September 2024 and January 2025 with five additional instruments (Figure \ref{fig: main_display}): X-shooter \citep{xshooter} on the VLT (program 114.2799.001, PIs: C.F. Manara, B. Tofflemire), the Las Cumbres Observatory (LCO) 1 m telescope network \citep{lco} with photometric monitoring in a suite of  filters (PI: B. Tofflemire), UVES \citep{uves} and MUSE \citep{muse} on the VLT (program 114.27MG.001, PI: E. T. Whelan), and GRAVITY \citep{gravity} on the VLTI (program 114.27PS.001, PIs: G. Bourdarot, M. Benisty). The LCO, X-shooter, and UVES data have already been partly published in \citet{Tofflemire25} and \citet{2025Hala}, and additional publications are in preparation to include these and the other datasets.

Coordination of this scale is rare and very difficult to achieve. Each proposal was independently submitted, approved, and executed by each observatory in planned coordination with the JWST-MIRI monitoring, which provided the motivation and backbone of the entire campaign. Coordination was only partly successful, as the JWST epochs were only scheduled one week in advance and most of the other instruments had their own scheduling constraints. An additional challenge came from JWST itself, which had to reschedule two epochs. Once in the first epoch originally scheduled on September 6th, which was simultaneous with the first iSHELL epoch and was expected to provide a baseline for comparison to conditions before the first accretion burst. This epoch was missed by JWST due to a failed guide-star acquisition. The second time was the third epoch originally scheduled on September 23rd in the quiescent phase before the second accretion burst, which would have provided an important reference point for the variability of MIRI spectra after the first strong burst. This second epoch was missed due to a problem with NASA’s Deep Space Network and the danger of data overflowing onboard JWST. The iSHELL epochs partly filled in the gaps left by JWST, by showing the CO line profile before the first burst and in between the first and the second bursts, supporting the interpretation of the MIRI data provided in this work.
Additional epochs from other instruments requested in queue mode on the VLT were missed due to weather conditions or scheduling constraint. Nonetheless, we were able to target the same 3 periastron passages and obtained a good coverage with most instruments (Figure \ref{fig: main_display} and Table \ref{tab: obs_campaign}). The only instrument that due to scheduling constraints ended up being scheduled in December without overlap with the MIRI monitoring is GRAVITY, except for a single epoch observed in September in between the first iSHELL and MIRI epochs. The rest of the GRAVITY epochs provided a dense sampling over one full orbit (7 epochs, from Dec. 22, 2024 to Jan. 7, 2025), which will be published in an upcoming paper. In the following, we describe in more details the observations included in this work from MIRI and iSHELL.

\subsection{JWST MIRI-MRS spectra} \label{sec: JWST_obs}
The mid-infrared spectra analyzed in this work were obtained with the James Webb Space Telescope \citep[JWST,][]{Gardner23} between September 11, 2024 and October 19, 2024 as part of Cycle 3 GO program 4727 (PI: A. Banzatti, Co-PIs: B. Tofflemire, Ch. Rab). DQ~Tau was observed with the Medium Resolution Spectrometer \citep[MRS,][]{jwst-mrs,Argyriou23} mode on MIRI \citep[][]{miri,miri2}. MIRI covers the full wavelength range of 4.9--28\,$\mu$m with resolving power of 1800-4000 \citep[][]{Argyriou23,pontoppidan24,banz25}. The spectra were extracted and wavelength-calibrated with the JDISCS pipeline as described in \cite{pontoppidan24}. For the data included in this work, we used the JDISCS reduction version 9 based on MRS pipeline version 12.1.5 \citep{MIRI_pip} and using the Calibration Reference Data System context jwst\_1364.pmap. The MIRI spectra were continuum-subtracted as explained in \cite{pontoppidan24} and \cite{banz25}, using the same parameters (window widths and excluded regions) for all epochs to minimize differences introduced by the continuum subtraction process. One full MIRI spectrum and its molecular emission is reported in Appendix \ref{app: additional_plots} for reference.
The MIRI continuum and representative gas emission lines of CO and \ce{H2O} are shown in Figure \ref{fig: main_display} as a function of time.

\begin{deluxetable}{l c c c c c c c c c c c c c c c}
\tabletypesize{\footnotesize}
\tablewidth{0pt}
\tablecaption{Epochs obtained in 2024 and analyzed in this work.}
\label{tab: miri_epochs}
\tablehead{ ID & Date & UT Time & $\rm{log}(L_{acc}/L_\odot)$ & $n_{13-26}$ } 
\tablecolumns{5}
\startdata
\multicolumn{5}{c}{JWST-MIRI} \\
B1.1 & Sep. 11$^{th}$ & 09:22:02 & 0.05 (0.09) &  0.06 \\
B1.4 & Sep. 14$^{th}$ & 00:55:43 & -1.17 (0.05) & -0.08 \\
B2.16 & Sep. 26$^{th}$ & 08:58:30 & -0.20 (0.39) & -0.13 \\
B2.21 & Oct. 1$^{st}$ & 11:22:22 & -1.10 (0.05)  & -0.17 \\
B2.26 & Oct. 5$^{th}$ & 23:06:01 & -1.33 (0.04)  & -0.22 \\
B2.27 & Oct. 7$^{th}$ & 02:16:06 & -1.36 (0.04) & -0.22 \\
B3.32 & Oct. 11$^{th}$ & 17:25:39 & -0.40 (0.07) & -0.19 \\
B3.34 & Oct. 13$^{th}$ & 19:47:06 & -1.25 (0.04) & -0.20 \\
B3.40 & Oct. 19$^{th}$ & 19:20:11 & -1.20 (0.14) & -0.21 \\
\hline
ID & Date & UT Time & $\rm{log}(L_{acc}/L_\odot)$ & 5~$\mu$m cont. (Jy) \\
\hline
\multicolumn{5}{c}{IRTF-iSHELL} \\
B1.-4 & Sep. 6$^{th}$ & 15:03:30 & -1.29 (0.1) & 0.39 (0.05) \\
B1.1 & Sep. 11$^{th}$ & 14:59:31 & -0.42 (0.1) & 0.41 (0.05) \\
B1.5 & Sep. 15$^{th}$ & 15:08:50 & -1.09 (0.1) & 0.44 (0.07) \\
B1.13 & Sep. 23$^{th}$ & 14:58:16 & 0.14 (0.1) & 0.53 (0.07) \\
B2.16 & Sep. 26$^{th}$ & 15:59:28 & -0.22 (0.1) & 0.49 (0.05) \\
B2.24 & Oct. 4$^{th}$ & 14:53:34 & -1.35 (0.1) & 0.36 (0.04) \\
B2.28 & Oct. 8$^{th}$ & 14:48:19 & -1.32 (0.1) & 0.31 (0.03) \\
B3.32 & Oct. 12$^{th}$ & 14:52:36 & -0.62 (0.1) & 0.40 (0.03) \\
B3.37 & Oct. 17$^{th}$ & 14:50:55 & -1.25 (0.1) & 0.36 (0.04) \\
\enddata
\tablecomments{See Figure \ref{fig: main_display} for the definition of epochs IDs. The UT time shows the start of each epoch. The 5~$\mu$m continuum flux and error for iSHELL epochs is the median and standard deviation flux at 4.9--5~$\mu$m (where iSHELL and MIRI overlap in wavelength). }
\end{deluxetable}

\subsection{IRTF iSHELL spectra} \label{sec: iSHELL_obs}
DQ~Tau was monitored with iSHELL at the IRTF \citep{ishell,ishell22} between September 6 and October 17, 2024 (PIs: B. Tofflemire, A. Banzatti). All epochs were requested to be within 1 day of a MIRI epoch, although the planning windows did not allow the final scheduling to achieve that request. The first and fourth iSHELL epochs were supposed to be near-simultaneous with MIRI, but JWST missed those epochs due to the aforementioned issues, and they had to be rescheduled at different times. Still, each iSHELL epoch ended up being $\le$ 3 days apart from each MIRI epoch. The iSHELL spectra were obtained with the 0".75 slit providing a resolving power R $\sim 60,000$ (corresponding to a velocity of 5~km/s) and were reduced and analyzed as described in \cite{banz22}. Flux calibration was performed with the standard procedure in Spextool v5.0.3 \citep{spextool} using the standard star HR~1165 observed with DQ~Tau in each epoch. Each spectrum covered 4.52--5.24~$\mu$m, i.e., overlapping with MIRI-MRS at 4.9--5.24~$\mu$m and revealing the full line kinematics of high-$J$ lines of CO (v1-0) between P25 and P45 observed at lower R with MIRI. Figure \ref{fig: main_display} includes the stacked profiles for the $v=1-0$ and $v=2-1$ lines.

\begin{figure}
\centering
\includegraphics[width=0.5\textwidth]{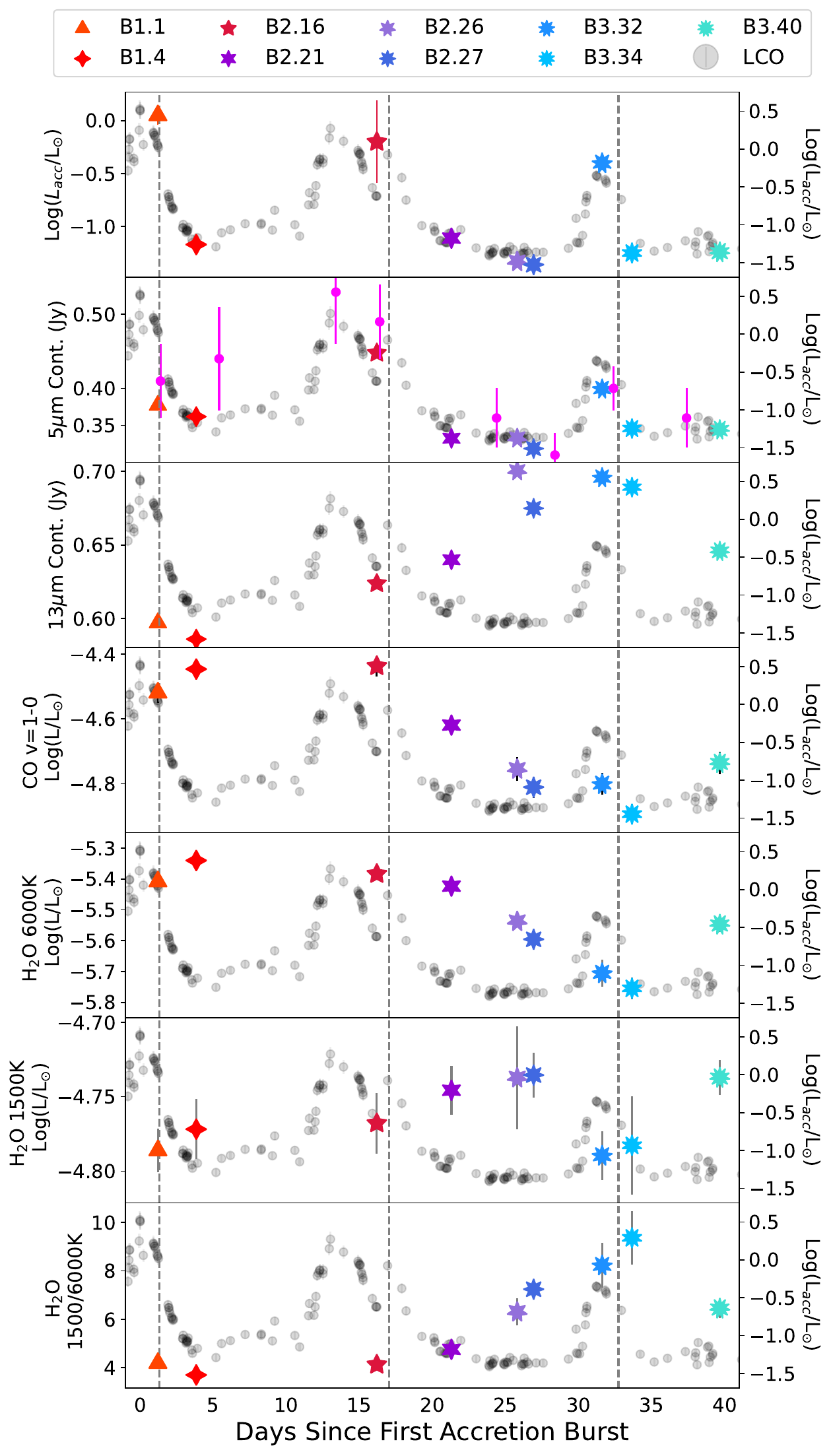}
\caption{ Light curves of the HI-derived accretion luminosity (Appendix \ref{app: updated_Lacc_relations}), 5$\mu$m and 13$\mu$m dust continuum flux, and different molecular lines tracing gas at different temperatures \citep[from hotter gas with the 6000~K lines to colder gas with the 1500~K lines, see][]{banz25}. The vertical dashed lines indicate the three periastron orbital phases. The accretion luminosity light curve measured with LCO \cite[from][]{Tofflemire25} is reported in grey using the second y axis to the right. The colored (red to blue) markers indicate the JWST-MIRI data. The magenta points in the 5~$\mu$m continuum panel show the fluxes measured in iSHELL spectra and reported in Table \ref{tab: miri_epochs}.}
\label{fig: line_var}
\end{figure}

\section{Analysis \& Results} \label{sec: results}
\subsection{Gas and Dust Emission Light Curves}
As a first step in this work, we analyze the measured variability of gas and dust emission tracers as a function of time, to investigate their potential connections to the accretion variability.
The line fluxes of different molecules observed in MIRI spectra are extracted from each epoch of DQ~Tau data using the same process described in \citet{banz25} as implemented in iSLAT \citep{iSLAT}, which is applied to the larger line list included in \citet{mallaney26}. We also measure the line-free continuum flux at 5, 13, and 26~$\mu$m in each epoch (from the continuum estimated as described above). In Figure \ref{fig: line_var}, the continuum flux and molecular gas line luminosity from each epoch are plotted in a time series as a light curve in reference to the X-Shooter-calibrated LCO accretion luminosity from \citet{Tofflemire25}. 

The 5~$\mu$m continuum flux is sensitive to the hotter dust that MIRI observes near and around the inner disk rim \citep[e.g. Figure 12 in ][]{Woitke2024}. In our observations, we find that it tightly follows the accretion pulses with the exception of the first epoch\footnote{We note that the 26 $\mu$m continuum instead is very luminous in the first epoch, see Appendix \ref{app: additional_plots}}. In Figure \ref{fig: line_var} we also include the 5~$\mu$m continuum flux as measured in the iSHELL spectra, which is more uncertain due to variable observing conditions and flux loss through the slit. Overall, the iSHELL continuum is consistent with the MIRI continuum within the errors, demonstrating the high quality of the ground-based observations obtained in this program. The MIRI 13~$\mu$m continuum flux (and similarly the 26~$\mu$m, included in Appendix \ref{app: additional_plots}), which trace colder disk regions at larger radii, instead show a very different trend: the flux increases over time and then decreases in the last epoch without being modulated by the individual bursts. The different behavior of dust continuum at different wavelengths already suggests that the inner disk reservoirs of dust and gas may be affected differently at different radii by the accretion process and variability in DQ~Tau, with a closer-in region more tightly connected to accretion and an outer region that varies on a different, longer timescale. This will be discussed by putting together all results in Section \ref{sec: discussion}.

\begin{figure*}
\centering
\includegraphics[width=\textwidth]{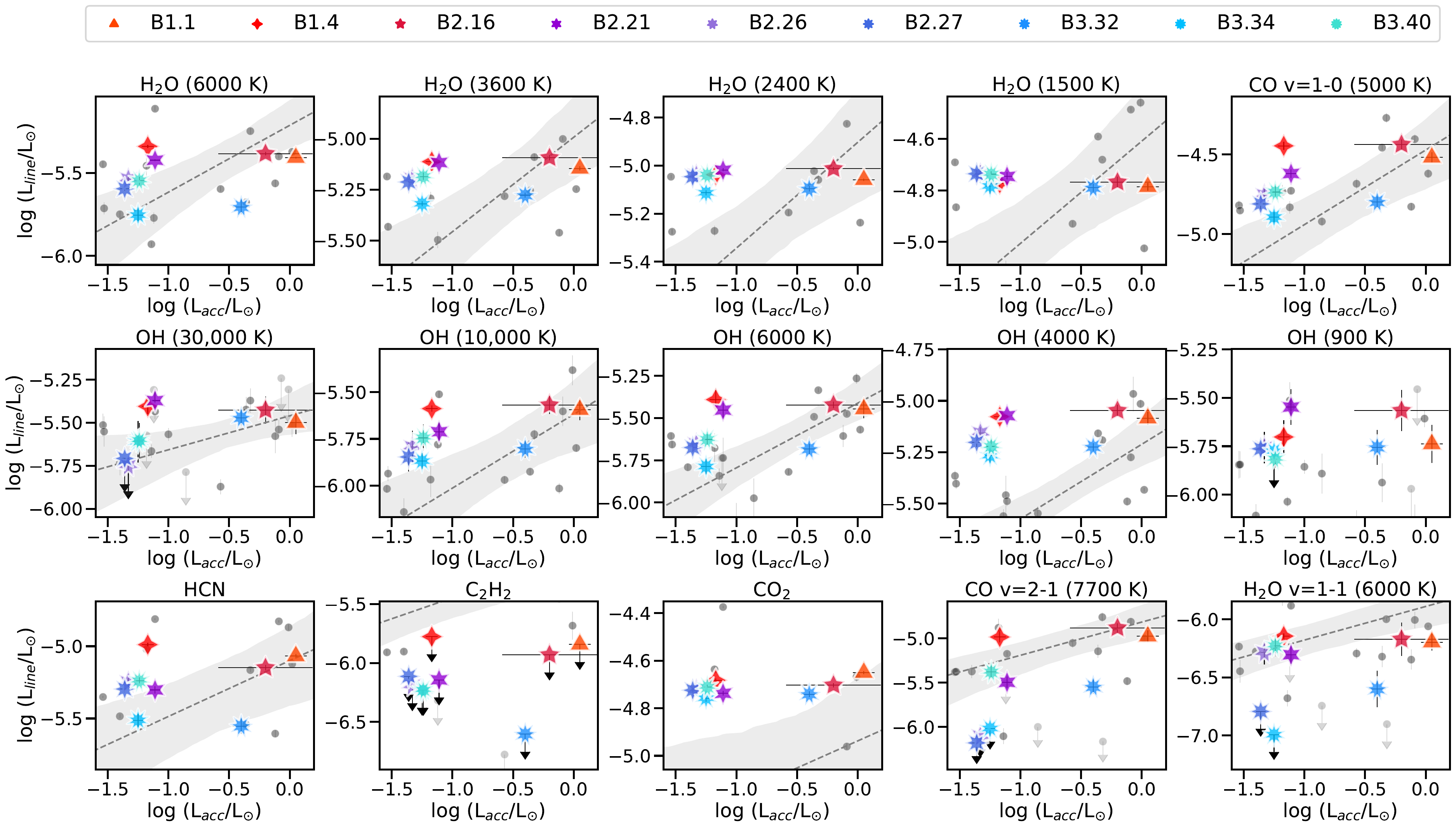}
\caption{Line luminosities for multiple molecules and energy levels (reported in parentheses) as a function of accretion luminosity, following the same grid and using the same JDISCS-C1 sample for reference as published in \citet{mallaney26}, along with its linear fits and 95\% confidence intervals. While the full correlations for the JDISCS-C1 sample are visible in \citet{mallaney26}, here we zoom in on the DQ~Tau datapoints (colored markers as in Figure \ref{fig: line_var}) to better visualize their distribution in reference to the JDISCS-C1 data and correlations (in grey).}
\label{fig: dqtau_correlations}
\end{figure*}

As for the molecular gas, we use emission lines from a range of $E_u$ \citep[which are indicated in Kelvin throughout this work, following those selected in][]{banz25} that trace hotter, warmer, and colder reservoirs to investigate variability at different disk radii. All the tracers of warm-hot gas, such as the higher-energy water lines, CO, OH, and HCN, show a global decrease in luminosity across the three burst covered in this campaign both in rotational and ro-vibrational lines (additional tracers are included in Appendix \ref{app: additional_plots}). Surprisingly, this global luminosity decrease is not modulated by the individual accretion bursts. The hotter gas tracers decrease to a minimum at about 33~days after the first burst and increase in the very last epoch, showing the exact opposite trend than the dust continuum at 13~$\mu$m. 
The 1500~K water lines, which are more sensitive to colder gas (temperatures of $<400$~K), exhibit overall a lower variability with an increase after 20~days followed by a decrease at 33~days and then increase in the last epoch as the hotter tracers. 

The 1500/6000~K ratio, the flux ratio of diagnostic water lines at these upper level energies as defined in \citet{banz25}, reflects the relative emission contribution from the colder reservoir in reference to the hotter reservoir and has been used in recent works as a proxy for the radial distribution of water in the inner disk. This line ratio in DQ~Tau follows a similar trend to that of the 13$\mu$m flux and inverse that of the 6000~K water lines, increasing after 20~days and then decreasing in the final epoch. More on the diagnostic ratios and radial gradient of water will be added in Section \ref{sec: water_diagrams}.

\subsection{Luminosity Trends with Accretion}
We analyze the measured gas variability as a function of the accretion luminosity, to test if accretion variability directly drives molecular emission variability following the global correlations defined by single-epochs spectra from a larger disk sample in previous work.
In Figure \ref{fig: dqtau_correlations}, we show molecular line emission as a function of accretion luminosity in DQ~Tau in comparison to correlations measured from a larger sample of disks around single T~Tauri stars from the JDISC Survey \citep{Arulanantham25} as presented in \citet{mallaney26} (i.e., excluding Herbig disks; this reference sample is called JDISCS-C1 in this work). Putting DQ~Tau, a close binary, in the context of disks around single stars is interesting because its molecular spectrum (as reported in this and previous works) is otherwise very similar to that of disks of single T-Tauri stars, and it is to date unclear how much the inner binary may or not affect the molecular reservoirs.

The molecular tracers included in Figure \ref{fig: dqtau_correlations} follow the same grid used in \citet{mallaney26} as based on the line list defined in \cite{banz25} and \citet{mallaney26}. For the accretion luminosity, we use values obtained from X-shooter and LCO as described in \citet{Tofflemire25} for DQ~Tau, and literature values for the other disks. 
We find that the luminosity of all molecular lines in DQ~Tau shows a similar range and scatter of values as in the larger JDISCS-C1 sample but generally does not follow tightly the global correlations. In at least one case, the \ce{H2O} 1500~K lines, the trend is not only different but opposite, with lines that increase in luminosity when \lacc is lower \citep[similar to the trend expected by thermo-chemical models in][]{Vlasblom25}. The DQ~Tau values seem to generally be most consistent with the trends from the larger sample in the three epochs during the bursts, and tend to be super-luminous in the lower-accretion epochs. This is particularly visible in the lower-energy water lines, in the 4000~K OH lines, and in \ce{CO2}. We remark that also the luminosity of [NeII] presents a similar trend as \ce{CO2} in being super-luminous and flat with change in \lacc (Appendix \ref{app: additional_plots}). On the contrary, \ce{C2H2} is only tentatively detected\footnote{There is a hint for \ce{C2H2} emission at 13.7~$\mu$m but it overlaps with multiple water and HCN transitions that make difficult to confirm its detection.} in DQ~Tau and it is definitely sub-luminous in comparison to the JDISCS-C1 sample. The similarities and differences between the molecular emission in DQ~Tau and in disks of single stars are very instructive on the effects of an inner binary on the inner molecular reservoir and will be later discussed together with the other results in Section \ref{sec: discussion}.

\begin{figure}
\centering
\includegraphics[width=0.4\textwidth]{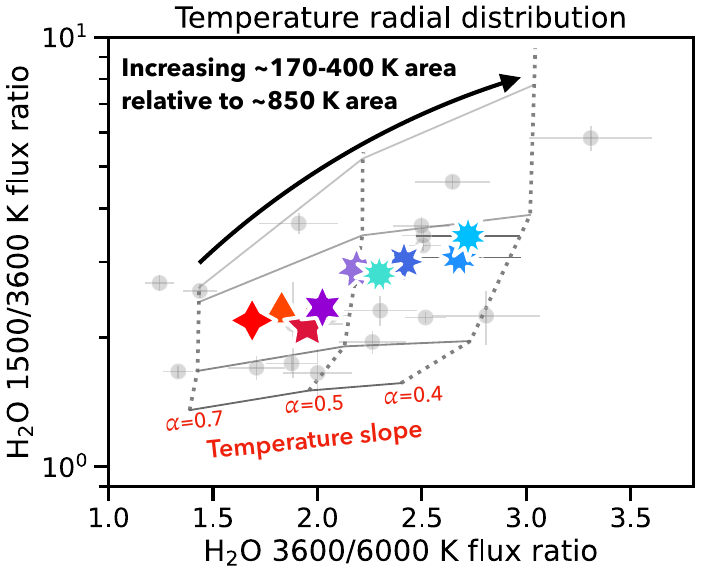}
\includegraphics[width=0.4\textwidth]{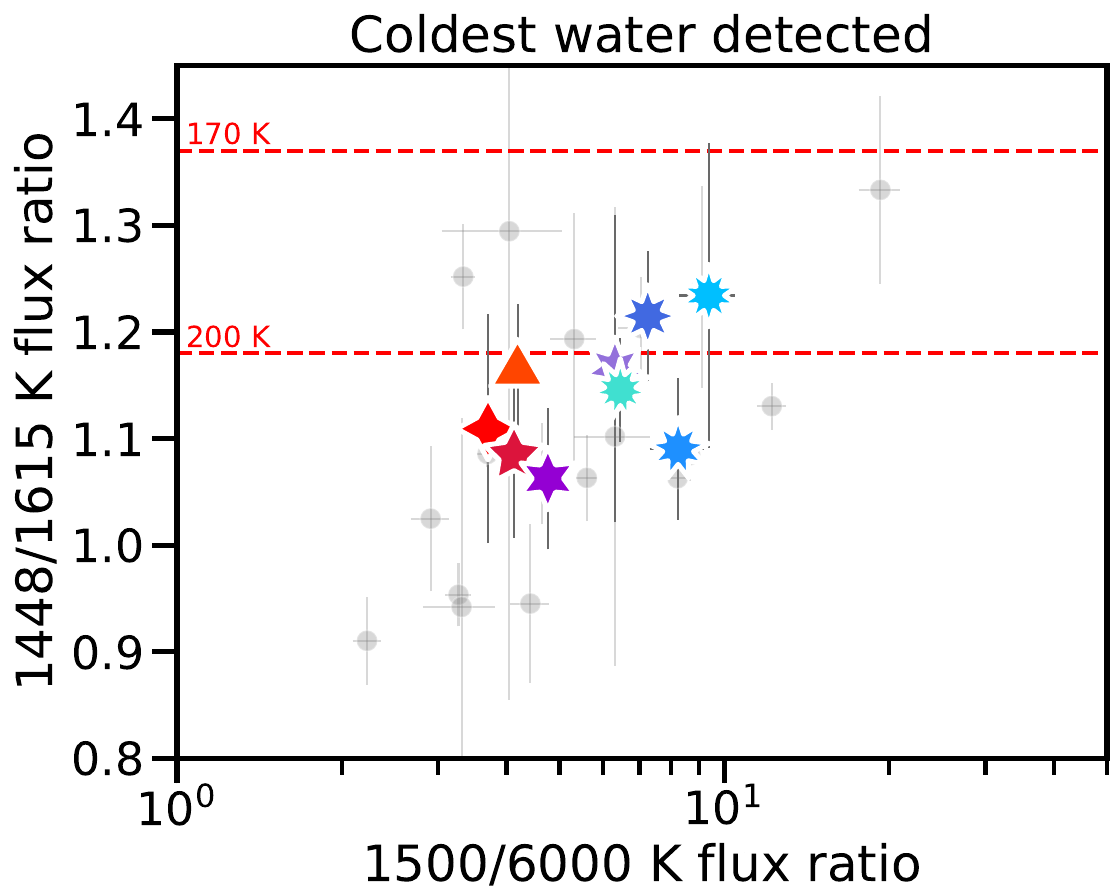}
\caption{Water diagnostic diagrams as defined in \cite{banz25} using flux ratios of lines with different upper-level energies as shown on the axes. Top: temperature gradient diagnostic comparing the variability observed in DQ~Tau in this program to the distribution of JDISCS-C1 targets. Bottom: Cold water diagnostic diagram comparing the variability observed in DQ~Tau in this program to the distribution of JDISCS-C1 targets from \cite{banz25}. The two red dashed lines represent the 170~K and 200~K temperatures, assuming a column density of $10^{17}$~cm$^{-2}$. The 1448/1615~K line flux ratio increases inversely with temperature (see Figure 11 in \cite{banz25}).}
\label{fig: diagnostic_diagrams}
\end{figure}

\subsection{Evolution in Water Excitation} \label{sec: water_diagrams}
We investigate the evolution in spectral properties of water with time to test if the radial temperature gradient changes in DQ~Tau and compare it with the diverse water distributions found in a larger disk sample in recent works.
Figure \ref{fig: diagnostic_diagrams} shows the main water diagnostic diagram defined in \citet{banz25}, using line ratios at different upper-level energies that are sensitive to water emission at different temperatures. The position of individual disks in these diagrams illustrates empirically whether they are dominated by a hotter inner reservoir or rather have stronger emission from colder water, by being, respectively, in the lower-left rather than upper region of the first diagram in the figure. Each disk can also be interpreted in this diagram in the context of multi-component slab models or a radial gradient. The grid of model tracks overlaid to the data in Figure \ref{fig: diagnostic_diagrams} (upper panel) assume power-law profiles in temperature and column density as $T = T_0 ( r / 0.5 \text{au})^{-\alpha}$ and $N = N_0 ( r / 0.5 \text{au})^{-\beta}$ as in \citet{munozromero24a} with fixed values as in \citet{banz25}: $T_0 = 450 K$, $N_0 = 1 \times 10^{18}$~cm$^{-2}$, $\beta = 1$, while $\alpha$ is varied between 0.4 and 0.7 and a cutoff radius $R_{\rm{cut}}$ between 0.75~au and 10~au \citep[see Figure 22 in][]{banz25}. A compact hot reservoir in this case, would have an $\alpha$ of 0.7 and minimum $R_{\rm{cut}}$ of 0.75, placing a given disk close to the lower-left corner of the model grid \citep[as it is the case for the disk of CI~Tau highlighted in][]{banz25}. DQ~Tau is close to that point in the first three epochs, demonstrating to be dominated by a hotter inner water reservoir.

The time-evolution of the water spectrum observed in DQ~Tau is interesting in the broader context of the values covered by the broader disk sample from JDISCS-C1 (shown in grey in this figure). DQ~Tau moves diagonally across most of the JDISCS-C1 sample from the bottom-left corner towards the upper-right corner of the model grid due to the decrease in hot water luminosity and the small increase in the cold water in some later epochs (see Figure \ref{fig: line_var}).
The line ratios of lower-energy vs higher-energy lines (3600K/6000~K and 1500/3600~K) therefore increase over time, indicating a flatter temperature gradient at later epochs, mostly driven by the decrease in the hot component. Recent works discussed these ratios in cases of increased cold emission at larger disk radii \citep{banz25,2025smith,Vlasblom25}, while in the case of DQ~Tau it is the decrease in the hot reservoir that drives most of the changes in line ratios.
However, the 1448/1615 K line ratio, which is sensitive to the coldest water detected with MIRI, clusters around values of $\sim 200$~K later in the monitoring, indicating increased emission from colder water near the snowline in support of the small increase in line emission from these lines already presented for the 1500 K lines in Figure \ref{fig: line_var}.

\begin{figure*}
\centering
\includegraphics[width=\textwidth]{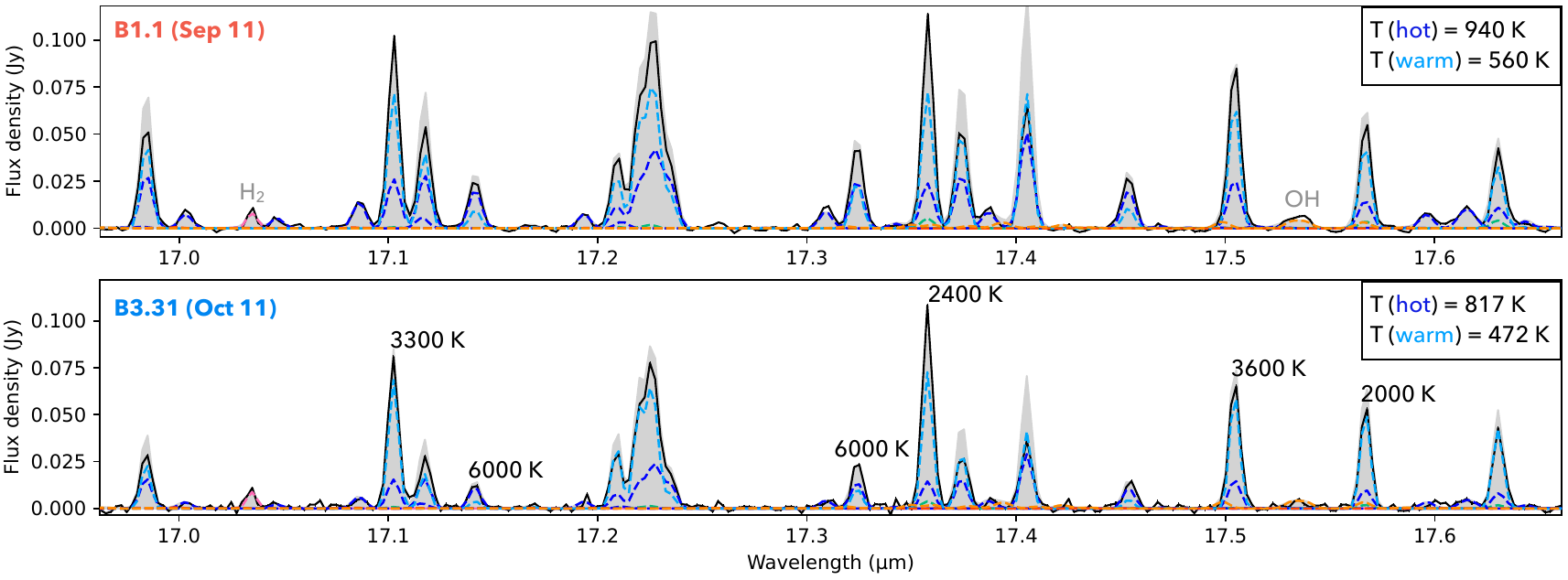}
\caption{Examples of best-fit slab models obtained from fitting the list of isolated \ce{H2O} rotational lines provided in \citet{banz25}, which are labeled in this figure with their approximate upper level energy for reference. We include the two epochs closest to the first and third accretion bursts to demonstrate the change in temperature in the hot and warm components. The cold component is prominent only in lines at longer wavelengths, not covered in this plot. The observed JWST-MIRI spectra are shown in black, the models in colors as indicated (with OH in orange and \ce{H2} in pink and total slab model flux in grey). The complete slab model fit results are reported in the Appendix. The models include line opacity overlap as implemented in iSLAT v5 \citep{iSLAT_codev5}.}
\label{fig: bestfit_examples}
\end{figure*}

\begin{figure*}
\centering
\includegraphics[width=\textwidth]{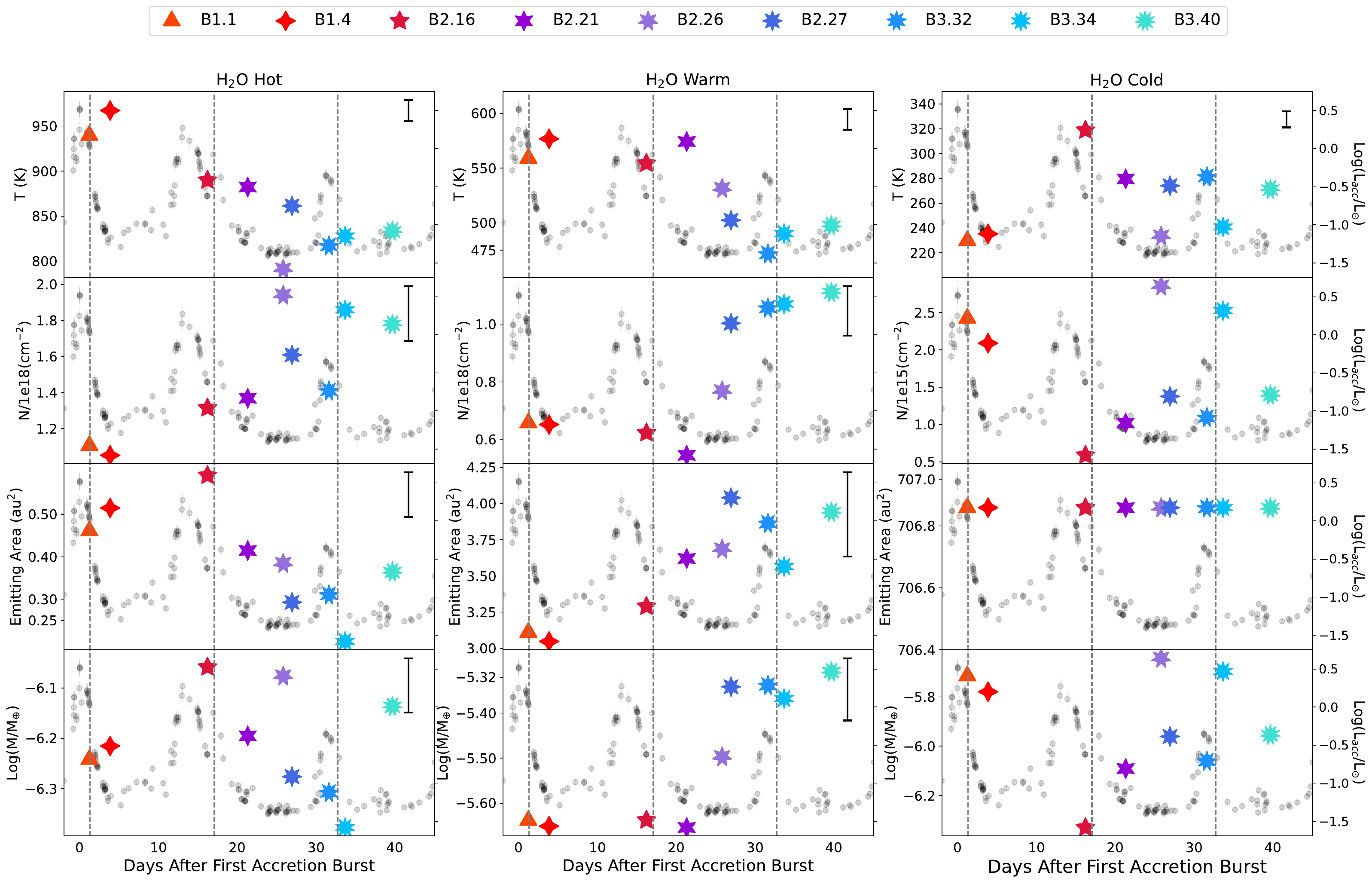}
\caption{Best-fit 3-component slab model values for temperature, column density, area, and mass as obtained from fitting the list of isolated \ce{H2O} rotational lines provided in \citet{banz25}, in reference to the accretion variability (grey datapoints in the background) as in Figure \ref{fig: line_var}. The complete slab model fit results are reported in the Appendix. The bar in the upper right of each panel is the median error across all epochs for each respective parameter. The column density and emitting area for the cold component are degenerate and converge to the upper boundary in emitting area.}
\label{fig: BestTNRM}
\end{figure*}

\subsection{3-temperature slab fits to water spectra} \label{sec: slab_fits}
We fit water spectra with a slab model to estimate quantitative changes in temperature, column density, and emitting mass as a function of time in DQ~Tau.
Following previous works that found evidence for radial gradients in water spectra observed in protoplanetary disks \citep[e.g.][]{banz23,munozromero24b,temmink25}, we perform 3-temperature slab fits to water emission in each epoch by fitting the list of isolated \ce{H2O} rotational $v=0-0$ lines at 10--27~$\mu$m provided in Tables 5 and 6 in \citet{banz25}, which are included in iSLAT. The three water components are fitted simultaneously to account for different contributions of each component to the different lines across energy levels, and the slab model adopted here is the same as in \citet{banz25} and similar to what used in other works, where the emission from each slab is described by an excitation temperature $T$, a column density $N$, and an emitting area $A$ \citep[for details, see e.g.][]{banz12,spectools_ir,iris}. We use the new HITRAN 2024 release \citep{HITRAN24} complemented with the HITEMP 2010 line list \citep{HITEMP10} for higher energy levels, which is necessary to model several lines at the shorter MIRI wavelengths \citep{banz25}. We assume an intrinsic broadening of FWHM = 1~km/s for all components. 

The results are reported in Figures \ref{fig: bestfit_examples} and \ref{fig: BestTNRM} and in Table \ref{tab: slabfits}. We find temperatures of $\sim 800$--1000~K, $\sim 450$--600~K, and $\sim 200$--320~K for the hot, warm, and cold components respectively, similarly to previous results from fits to water emission observed with MIRI in samples of T~Tauri disks \citep[e.g.][]{banz23b,munozromero24b,temmink25}. The column densities are also similar to previous results ($\sim 10^{18}$~cm$^{-2}$), indicating optically thick emission in the warm-hot reservoirs, and lower in the cold reservoir \citep{banz25}. The cold component is degenerate in column density and emitting area, providing only an estimate of the observed water mass as $\sim 10^{-6}$~M$_{\oplus}$, in between the masses estimated for the hot (lower) and warm components (higher). 

Figure \ref{fig: BestTNRM} shows the best-fit results as a function of time during this monitoring campaign. The hot component globally decreases in temperature and emitting area and increases in column density as a function of time across the entire monitoring, i.e., with a longer timescale than the individual bursts. The trends in temperature and column density show very small scatter, with only one epoch (B2.26, from Oct. 5) that shows a sudden larger decrease in temperature and increase in column density. Such a clean trend demonstrates the quality of the slab model results as based on MIRI spectra (see more in Appendix \ref{app: additional_plots}). 
The warm component also decreases in temperature with time, but with a more sudden change after epoch B2.21 (Oct. 1) that is visible in the sudden increase in column density. Contrary to the hot component, the emitting area increases with time. 
The cold component is more uncertain due to the fewer lines that are sensitive to it, and it shows a scatter of temperatures and masses without a clear trend with time. The emitting area for the cold component reaches the upper boundary of our slab models and values are all very similar in Figure \ref{fig: BestTNRM}.

\begin{figure}
\centering
\includegraphics[width=0.45\textwidth]{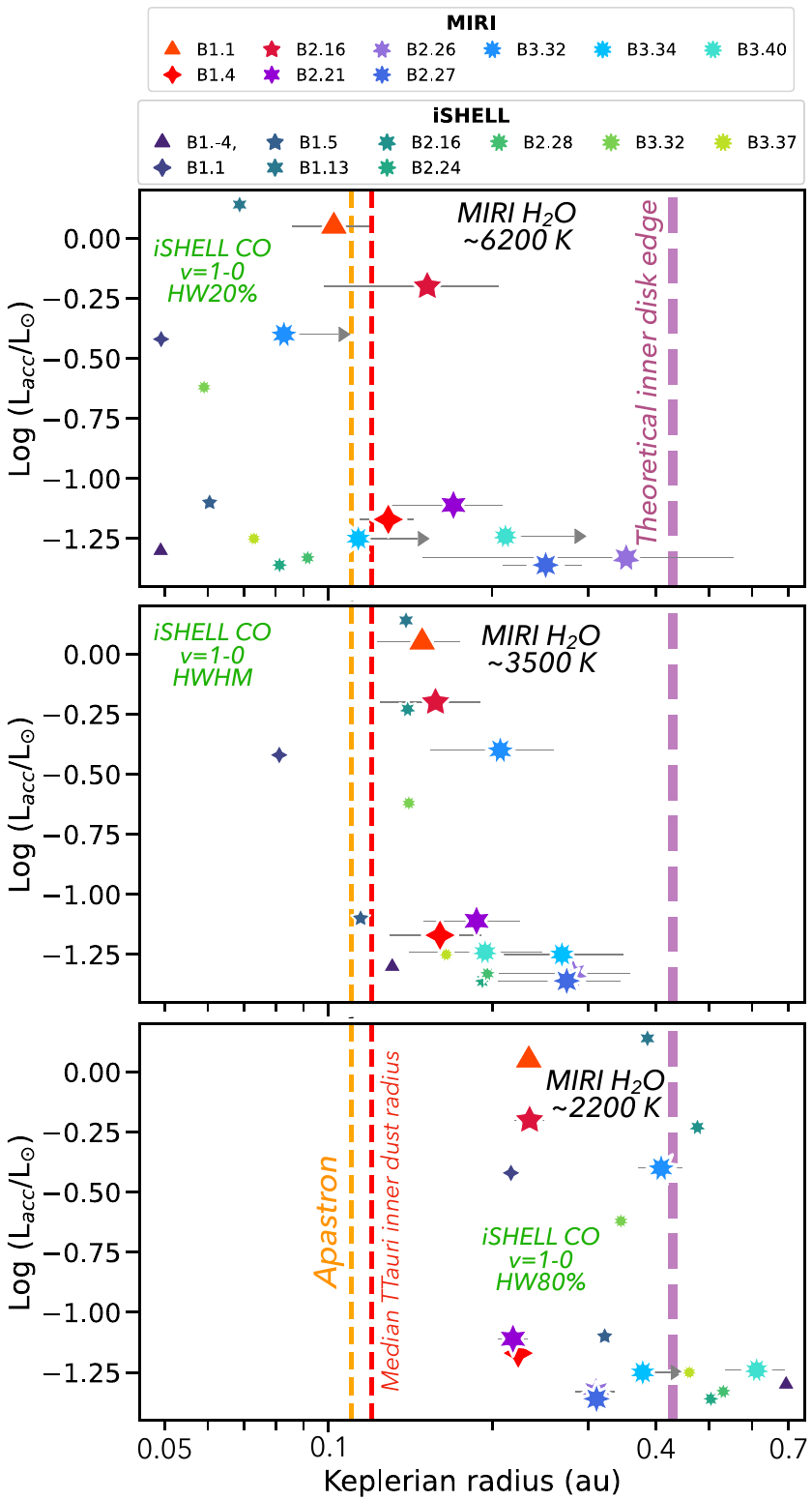}
\caption{ Time-resolved Doppler mapping of water lines from the line broadening measured in MIRI spectra (colored datapoints, same as in previous figures), compared to Keplerian radii measured from fully-resolved CO line profiles measured from iSHELL spectra (green datapoints from darker to lighter with time). Emission from CO and water is observed to extend well within the region in between the binary orbit, and to move out with time. The red dashed lines shows the median inner dust rim radius measured in single T~Tauri stars with stellar luminosity similar to the total luminosity of the DQ~Tau binary, taken from Figure 21 in \citet{banz22}. The apastron distance and theoretical inner disk edge are from Table \ref{tab: DQTau props}. Error-bars in radii for the 2200~K water lines are smaller than the marker size in a few epochs.}
\label{fig: Kepl radii}
\end{figure}

\subsection{Keplerian emitting radii from Doppler broadening} \label{sec: res_radii}
We estimate the physical emitting radii of water in the inner disk to investigate how those vary as a function of time and accretion in DQ~Tau.
We follow the analysis presented in \cite{banz25} to estimate the Keplerian broadening of MIRI lines by de-convolving the measured line broadening with the MRS resolving power profile in each sub-band as published in \cite{pontoppidan24} and updated in \cite{banz25}. The Keplerian broadening corresponding to the half-width-at-half-maximum (HWHM = FWHM/2), which can be marginally resolved above the MIRI resolving power, is then converted into a characteristic Keplerian radius by using the known disk inclination and total stellar mass for the binary (Table \ref{tab: DQTau props}). This approximation was adopted in \citet{Carr01} and should be valid as long as the gas orbits both stars, which we assume to be true for the molecular lines studied in this work in absence of direct evidence for the contrary. We caution, however, that the inner region close to the binary in DQ~Tau is expected by models to be dynamically very complex and likely include asymmetric streams (see more in Section \ref{sec: discussion}) that could affect the observed line profiles in ways that are currently unknown, as it is unclear if and where molecular gas would survive in these streams (which may instead be purely atomic due to the strong irradiation near the stars). We therefore simply estimate radii from the molecular line widths assuming they are dominated by Keplerian motion around both stars, which should give a reliable radial measurement at least in a relative sense between epochs, if not an absolute one.

\begin{figure}
\centering
\includegraphics[width=0.47\textwidth]{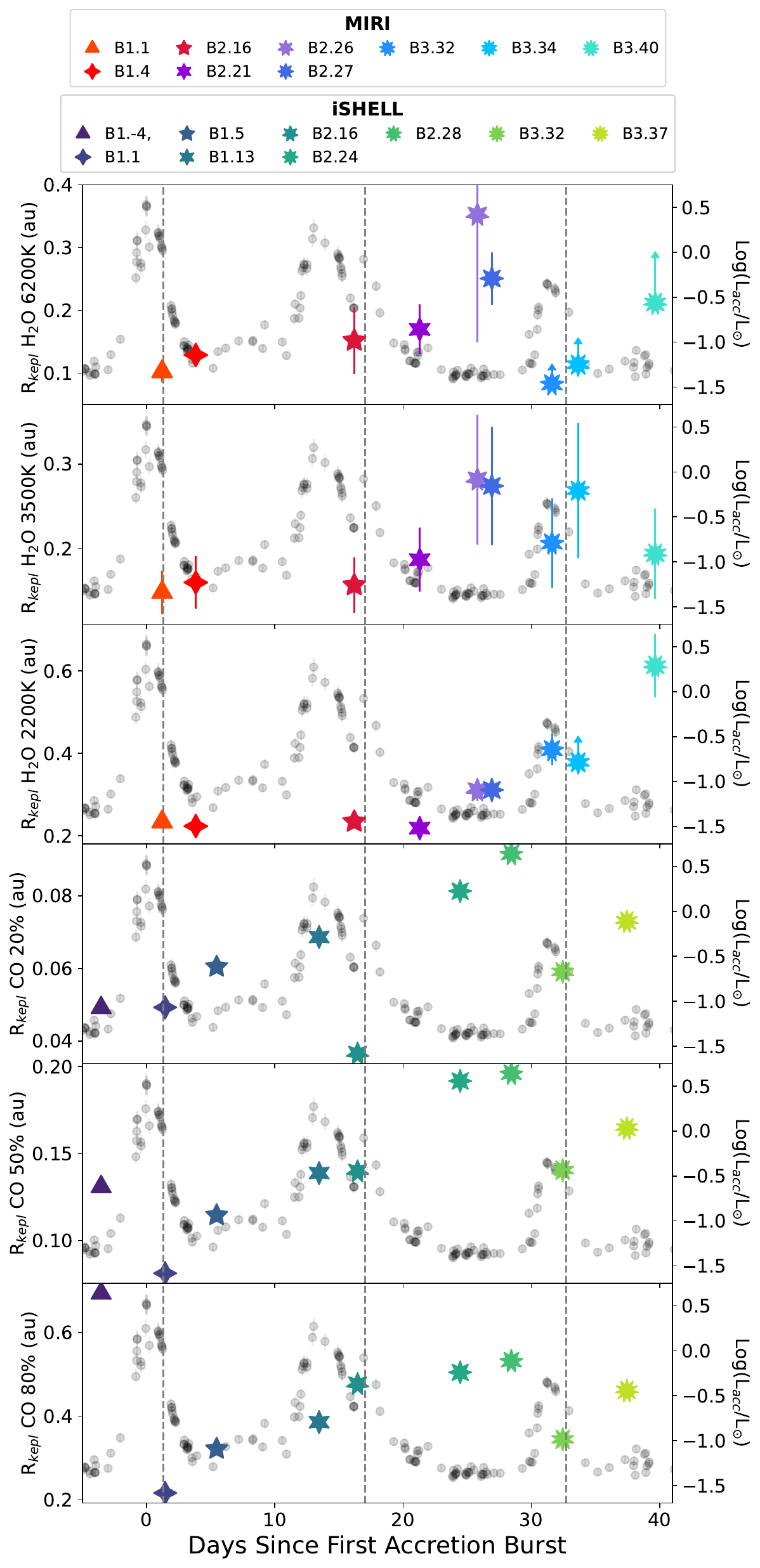}
\caption{Similar to Figure \ref{fig: line_var} but showing the Keplerian radii estimated from line broadening (see Section \ref{sec: res_radii}).}
\label{fig: Kepl radii vs time}
\end{figure}

In Figures \ref{fig: Kepl radii} and \ref{fig: Kepl radii vs time} we report the average Keplerian radius as a function of accretion luminosity and time for water lines at three values in upper level energy (which are sensitive to different regions along the radial gradient, see \citet{banz25} and Appendix \ref{app: additional_plots}); the complete Doppler maps for all lines and epochs are reported in Appendix \ref{app: additional_plots}. All rotational water lines observed with MIRI are broader in the first three epochs of this program, spanning the first two accretion bursts, and they recede to larger radii with time. The Keplerian radius of the $\sim 6200$~K lines, which are dominated by the hot slab component (see Figure \ref{fig: bestfit_examples}), is as small as $\sim 0.1$~au at the beginning of the monitoring, indicating that the hotter water reservoir extends to an inner region that overlaps with the binary separation at apastron. In the following epochs, the decreasing line broadening indicates that the hotter water emission recedes out to $\sim 0.3$~au in the lowest accretion epochs. In the last three epochs, the lines are unresolved and only provide lower limits. 

The $\sim 3500$~K lines, which are composed for $\sim 1/3$ in flux by the hot component and $\sim 2/3$ by the warm component (see Figure \ref{fig: bestfit_examples}), are resolved in all epochs and show first a recession out to $\sim 0.3$~au, coming back to $\sim 0.2$~au at the peak of the third burst, then out and in again.
The 2200~K lines, dominated by the warm water reservoir (see Figure \ref{fig: bestfit_examples}), show larger radii than the higher-energy lines in all epochs (from $\sim 0.2$~au earlier in the monitoring out to 0.4--0.6~au later), as expected from a radial gradient traced by water emission lines where the water temperature decreases with increasing distance from the star \citep{munozromero24b,temmink25,banz25}. The width of these lines narrows with time, showing that the emission moves to larger radii until the end of the monitoring. 
The lower-energy lines covered by MIRI with $E_u \sim$1500~K near 23.85~$\mu$m are not resolved in any epoch, placing a lower limit on their emission at $>0.2$~au at all times during this monitoring; these lines can be seen in Appendix \ref{app: additional_plots}.

As for the CO lines observed with iSHELL, their velocity profile is fully resolved at a spectral resolution of $\sim 5$~km/s. We use the stacked velocity profile from $v=1-0$ lines between $J = 25$ and $J = 31$, the least affected by telluric absorption and covered at the shortest MIRI-MRS wavelengths, following the procedure developed in \cite{banz22} and we measure their half-width line velocity at 20\%, 50\% (the HWHM), and 80\% of the line peak value to derive a characteristic Keplerian radius at these velocities. We include these three estimates in Figure \ref{fig: Kepl radii} for comparison to the radii from the HWHM (the only measurement that can be made from the limited velocity information provided by MIRI) of water lines at different energies. The CO radii from the 20\%, 50\%, and 80\% of the line peak span the entire radial region of the water lines across energy levels, supporting a similar emitting regions as found in previous work \citep[e.g. Figure 13 in][]{banz23}. The CO FWHM from iSHELL spectra is in the range 50--77~km/s, with the broadest lines during the strongest accretion burst on September 11 when iSHELL and MIRI observed DQ~Tau almost simultaneously (Figure \ref{fig: main_display}). 

Using system parameters from Table \ref{tab: DQTau props}, the CO $v=1-0$ HWHM measured with iSHELL during the MIRI monitoring translates into Keplerian radii of 0.08--0.14~au in September during the first two strong bursts and 0.14--0.2~au in October (see Figure \ref{fig: Kepl radii}). The line widths at 80\% of the peak are in the range 30--47~km/s corresponding to Keplerian radii of 0.2--0.7~au, indicating the CO emission extends at least out to these radii that overlap with the lower-energy water lines presented above. The CO line widths at 20\% are in the range of 70--100~km/s, corresponding to 0.05--0.09~au again with broader values (smaller radii) at the peak of the three bursts. The $v=2-1$ lines are stronger in September and match the high-velocity wings of the $v=1-0$ lines, indicating similar emitting radii down to at least $\sim 0.05$~au. The CO line widths at 20\% therefore demonstrate high-velocity gas well within the binary orbit (Figure \ref{fig: Kepl radii}), which would suggest that the high-velocity wings of CO lines may be affected by complex dynamical motion of the gas as stirred by the binary.

The CO radii globally move out with time with some modulation near periastron, where the CO line widths show a sudden increase suggesting the gas temporarily moves closer in, visible in the third burst at all velocities and also in the second burst in the FW20\% (Figure \ref{fig: Kepl radii vs time}). Overall, the radial emission of molecular gas therefore shows longer-term variations than the single orbital period, similarly to the longer-term trends found in the line luminosity (Figure \ref{fig: line_var}) and in the best-fit slab components (Figure \ref{fig: BestTNRM}).

\section{{Discussion}}\label{sec: discussion}
A first remarkable result of this monitoring is that it demonstrates the distinct variability that different tracers have within the MIRI spectra (Section \ref{sec: results} and Figure \ref{fig: line_var}). Some are tightly connected to the accretion variability estimated from LCO photometry and X-shooter spectroscopy (the HI lines and 5~$\mu$m continuum, both known to trace or be connected to accretion), some show a global decrease across multiple binary orbits (the hotter molecular gas emitting from the innermost disk region), while some show the opposite trend of increasing with time (the 13~$\mu$m and 26~$\mu$m continuum and the colder water tracers, which emit from larger disk radii).

The trends observed in luminosity and line widths in DQ~Tau show that the gas and dust emission variability observed from its inner disk reflects multiple effects happening over a range of disk radii, beyond just the excitation and chemical reactions from a variable UV irradiation in a static disk. In this section, we present our current best effort to reconcile into a unified picture the results from this campaign by taking into account models and interpretations that have been previously published. Without direct spatial information and months-to-years monitoring of molecular gas and dust tracers, parts of this picture remain speculative and are intended to provide the community with plausible scenarios to be tested and improved in future work.

\subsection{Dynamical rather than chemical changes} \label{sec: disc_singlestarstrends}
The tight correlations with accretion luminosity in T~Tauri disks \citep{banz20,banz25,mallaney26} suggest that inner disk irradiation is generally the major driver of the infrared molecular luminosity through an area effect (higher accretion drives a larger emitting area). Given that the inner molecular gas in DQ~Tau extends to an inner radius consistent with what measured in disks around single stars of similar luminosity \citep[0.05--0.3~au,][]{banz22,banz25}, it is reasonable to expect that its luminosity should follow similar correlations with \lacc, in absence of other effects.
Instead, DQ~Tau generally cuts across these correlations almost horizontally (Figure \ref{fig: dqtau_correlations}), and its decrease in emission from the hotter gas tracers does not simply follow \lacc. While thermo-chemical disk models of a generic T~Tauri disk explain the observed correlations between UV irradiation and water luminosity with gas-phase formation and excitation \citep{Woitke2024,Calahan26}, they also show that the chemical relaxation timescales in the infrared emitting region should be of the order of 0.1--10~yr \citep{Woitke2009,Woitke2024}, much longer than the accretion burst durations observed in DQ~Tau ($< 10$~days). According to these models, the molecular emission in DQ~Tau does not have time to reach equilibrium with the simultaneous UV irradiation \citep[see also][]{Kanwar25_CHequ}, which would therefore explain why its molecular emission is not following the irradiation variability from the individual accretion bursts.

Rather than on chemical changes caused by variable irradiation, our analysis shows that the molecular emission in DQ~Tau mostly depends on inner disk dynamics through a variable emitting region (Figure \ref{fig: Kepl radii}) and gas mass (Figure \ref{fig: BestTNRM}), which may be connected to the stellar accretion rate but not necessarily tightly correlated to it during short-term variability. The kinematic variability observed in the high-velocity CO gas in DQ~Tau is not uncommon in disks \citep[e.g. Figure 8 in][]{banz22}, suggesting that complex dynamics and a variable emitting region for the molecular gas may be more common than currently thought even in disks around single stars. Part of the scatter observed in the accretion luminosity correlations from the larger sample of T~Tauri disks could therefore also be due to dynamics rather than chemistry. Single-epoch MIRI spectra of other disks may offer only a snapshot at random phases within the cloud of datapoints that each disk could potentially show, if it were observed as a function of time. Our analysis of DQ~Tau suggests that it may be more important to determine the average accretion rate of each star and compare the MIRI molecular spectra to that, rather than to the specific value measured in a single epoch (whether from MIRI or available from the literature). Doing so could in principle reduce the scatter in the luminosity correlations like those presented in Figure \ref{fig: dqtau_correlations}; this is a test that could be done in future work on a stellar sample where accretion is monitored daily over several weeks.

\begin{figure*}
\centering
\includegraphics[width=1\textwidth]{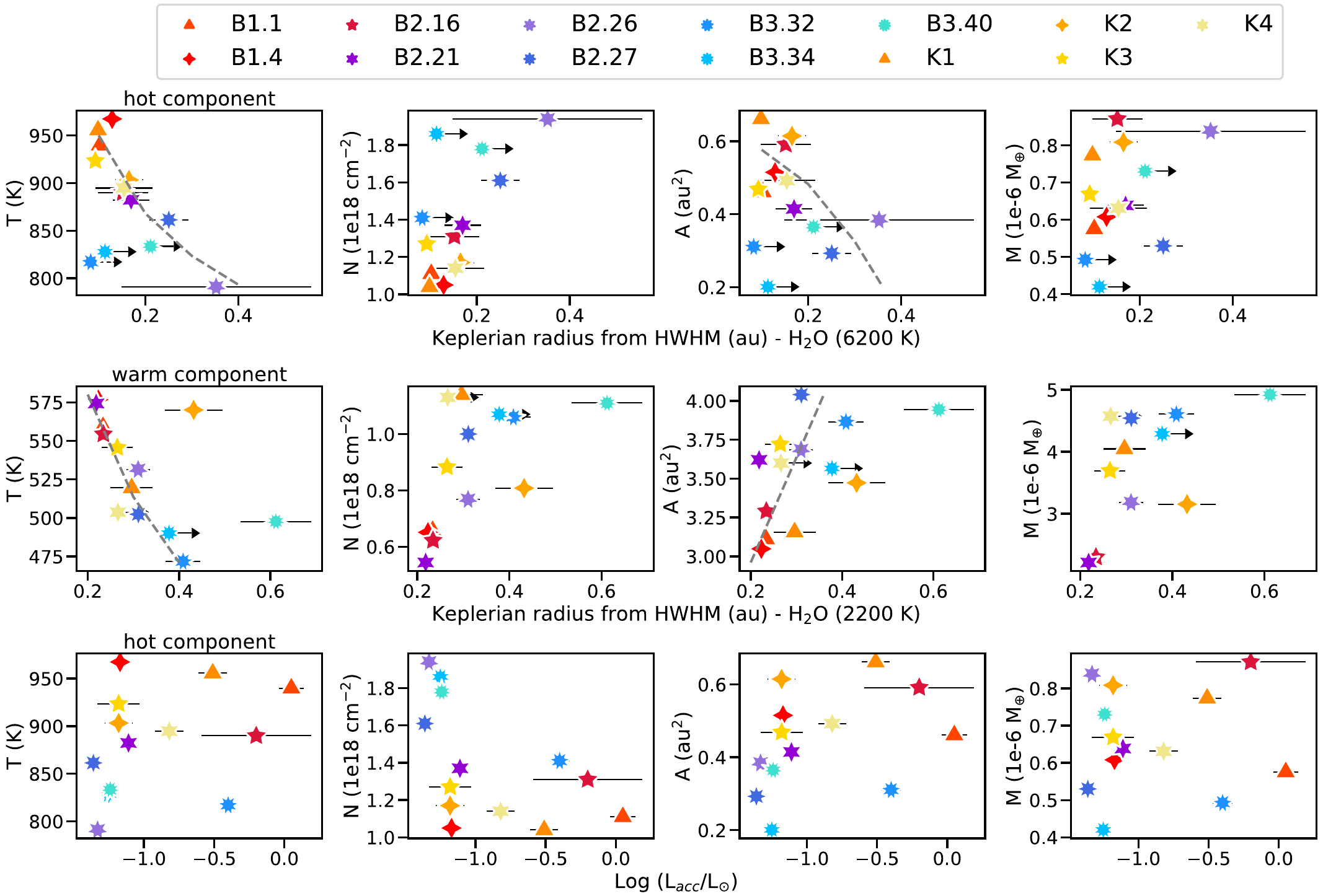}
\caption{Slab model fit results (from Figure \ref{fig: BestTNRM}) for the hot component (top) and warm component (middle) of water, shown as a function of the Keplerian radii independently measured from the line broadening (from Figure \ref{fig: Kepl radii}). The dashed curves in the temperature plots show power-law profiles with indices of -0.13 (hot component) and -0.3 (warm component), while the area plots show the expected trends from disk annuli that shrink (hot component) or expand with time (warm component), see text for details. The additional four epochs taken in 2026 (labeled K1 through 4) and published in \citet{kospal25} are included for comparison and agree with the values and trends found in our program. The plot in the bottom shows the hot component results as a function of \lacc; contrary to what reported in \citet{kospal25} based on 4 epochs, there is no correlation between the water temperature and \lacc.}
\label{fig: Slab_results_+radii}
\end{figure*}

\subsection{Variable gas within an inner disk cavity}
Dynamical models expect that the inner binary in DQ~Tau would produce a low-density inner cavity in the disk out to $\approx 0.4$~au (Section \ref{sec: intro} and Table \ref{tab: DQTau props}). Such a small cavity would not be detected in the infrared continuum slope index $n_{13-26}$ \citep[see its discussion in][]{DAlessio2006,furlan09,banz20,mallaney26}, which in fact is zero or slightly negative in all DQ~Tau epochs (Table \ref{tab: miri_epochs}). Moreover, this cavity is probably not fully devoid of dust, as dusty accretion streams may still feed the inner region as well as accretion onto the stars \citep{Jensen97,Carr01,Muzerolle19}. Nonetheless, spatially-unresolved dust-temperature fits to near-infrared dust continuum in DQ~Tau have suggested the presence of hot (1000--1600~K) circumbinary dust at an orbital radius of $\sim 0.2$~au \citep{Muzerolle19}, possibly moving in and out with time \citep{kospal25}. 

In this context, the Doppler broadening in molecular lines offers an independent estimate of the orbital radius for the emitting molecular gas. Assuming Keplerian motion, the ro-vibrational CO emission was found to extend inward to the binary orbit in \citet{Carr01}, which we support with our new observing campaigns with iSHELL (where double-peaked line profiles support the Keplerian interpretation, see Section \ref{sec: res_radii}). The novelty of the MIRI spectra is to provide radii estimates from water rotational lines \citep{banz25}, and in this work -- for the first time -- as a function of time in one disk. In Figures \ref{fig: line_var} and \ref{fig: Kepl radii}, we have shown that the water and CO lines move out to larger radii over the course of our program, with the higher-energy lines showing some modulation towards the end of the monitoring. 

We further explore this scenario in Figure \ref{fig: Slab_results_+radii}, where we combine the 3-component slab model results (Section \ref{sec: slab_fits}) to the $E_u$-averaged Keplerian radii from Figure \ref{fig: Kepl radii}. While the line broadening at different energies does not fully separate into the different best-fit slab components, the hot component dominates the flux and broadening of higher-energy lines and the warm component dominates the lower-energy lines (see examples in Figure \ref{fig: bestfit_examples}). In Figure \ref{fig: Slab_results_+radii} we therefore take the 6200~K line broadening as mostly reflecting the hot slab component and the 2200~K lines as mostly reflecting the warm component. The two axes in this figure combine independent and complementary information: the slab results only provide the equivalent emitting area of a disk annulus that is represented by a single temperature and column density, but they do not tell us where the ring is in the disk, while the line broadening is sensitive to the actual orbital radius of the emitting gas. The combination of this complementary information demonstrates that the temperature decrease found as a function of time in both the hot and warm components (Figure \ref{fig: BestTNRM}) happens while the gas emission is receding to larger disk radii. In other words, it is not that the inner disk gas is getting cooler due to a decrease in irradiation (which in fact increases again in each accretion burst): instead, the emitting gas is moving to larger radii and therefore tracing gas at lower temperature. The temperature profiles $T = T_0 ( r / R_{0})^{-\alpha}$ defined by the two components and their emitting radii are shown with dashed lines in Figure \ref{fig: Slab_results_+radii} and correspond approximately to the following values: $T_0 = 950~K$, $R_{0} = 0.1~$au, and $\alpha = 0.13$ for the hot component, and $T_0 = 580~K$, $R_{0} = 0.2~$au, and $\alpha = 0.3$ for the warm component.

The emission in the hot component shows a decrease in emitting area with time, consistent with a hot inner annulus that becomes narrower as the gas recedes to larger radii by losing the innermost high-velocity gas. The ground-based CO lines observed with iSHELL demonstrate that the high-velocity wings (as well as the $v=2-1$ lines) disappear at 28--32 days into the monitoring, when the high-energy water lines reach a minimum luminosity even through the third burst recorded in this program (Figures \ref{fig: main_display} and \ref{fig: line_var}). 
As a proof of concept, in Figure \ref{fig: Slab_results_+radii} we plot the area of an annulus where the outer radius is fixed at 0.44~au (the theoretical inner disk edge, Table \ref{tab: DQTau props}) and the inner radius varies as shown by the Keplerian radius of the water lines used in the x axis of the figure (between 0.1~au and 0.35~au). This serves just to demonstrate that, assuming the Keplerian radius from the line width traces close to the inner radius of the emitting annulus of gas, the recession of optically-thick molecular gas with time after the inner cavity is filled with gas in the first epochs can reproduce the decreasing trend in emitting area independently retrieved from the slab fits to the line fluxes.
In addition, the warm component increases in emitting area and mass as the emission moves to larger radii, as expected for an annulus of gas that becomes larger. If we take 0.45~au as the inner radius of this annulus, outside the outer radius of the annulus adopted for the hot component, the outer radius just needs to increase from 1.07~au to 1.22~au to reproduce the trend in emitting area found in the warm component. In this case, the measured line width is still significantly contaminated by the hot component (see Figure \ref{fig: bestfit_examples}) and would still be tracing smaller radii than the actual annular emission of the warm component.

\begin{figure*}
\centering
\includegraphics[width=\textwidth]{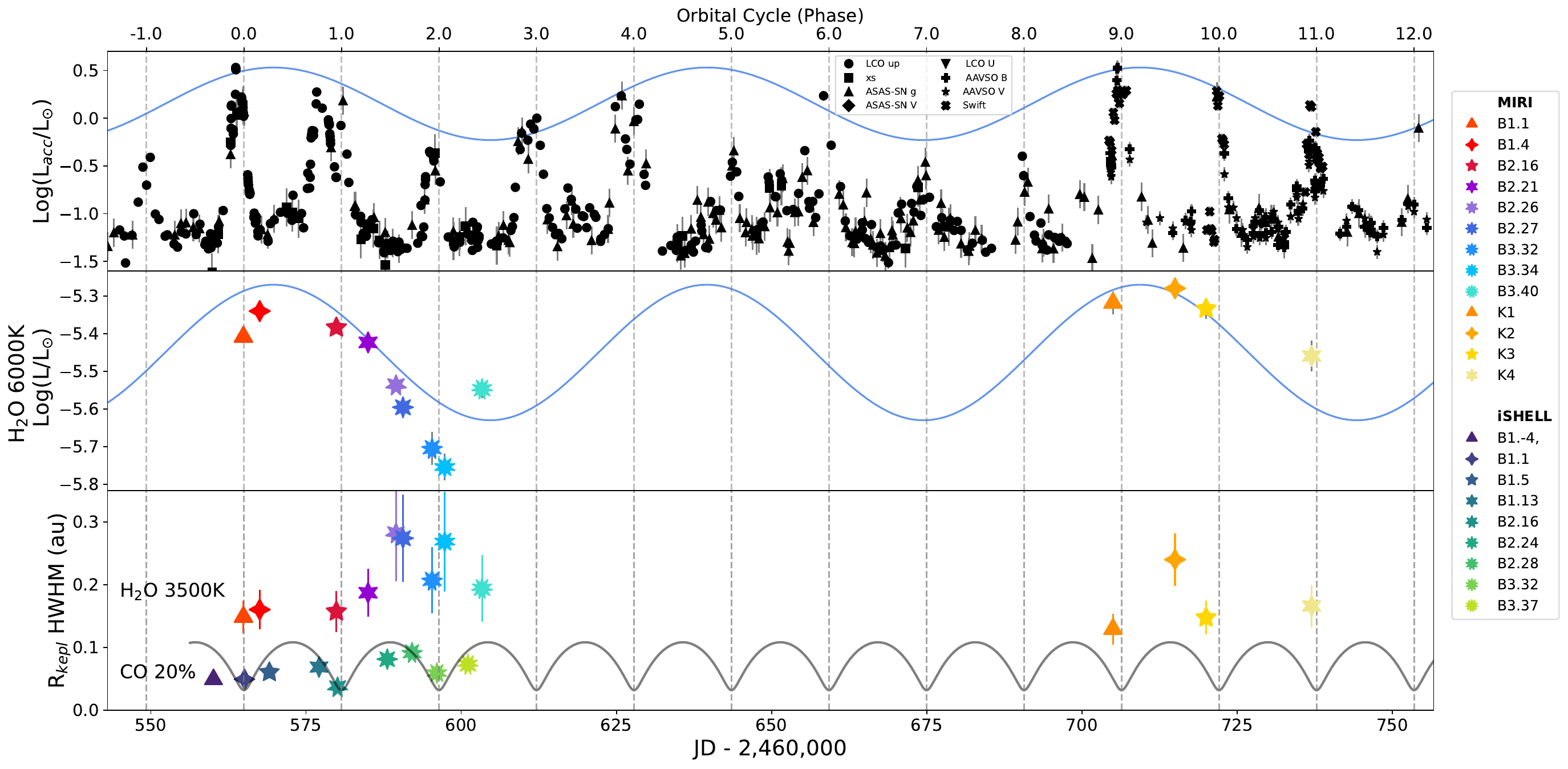}
\caption{Overview of our proposed interpretation to reconcile all the observables with a longer-term modulation of molecular gas near the inner disk edge during this monitoring campaign. The top panel displays the full \lacc time series collected in our program and in \citet{kospal25}, along with a sinusoidal curve to mark the tentative longer-term periodicity ($\sim 5 P_{bin}$) in the burst peak amplitude discussed in the text. The middle and lower panels show, as representative examples from the analysis in this work, the 6000~K water luminosity and the Keplerian radii for the FW20\% CO and 3500~K water lines, including for the later epochs from \citet{kospal25}. The barycenter distance of the primary is shown in the bottom plot for reference to the gas emission radii, illustrating how the high-velocity CO gas is temporarily pulled inward at the second and third periastron.} 
\label{fig: disc+cartoon}
\end{figure*}

In summary, the results from the slab model fits and the radii from the line broadening, which are completely independent in this analysis, are consistent with a common scenario where the inner disk cavity produced by the binary in DQ~Tau is filled with hot molecular gas at the beginning of our monitoring campaign, when the first two strong accretion bursts are observed. After that, a decrease of the molecular gas within the inner cavity is consistent with the decrease in line luminosity and excitation temperature and narrower line profiles that become dominated by colder gas at larger radii possibly due to a combination of disk draining by accretion and gas sweeping and outward pushing by the binary interaction. The similar column density we measure in the hot--warm slab components suggests that we are following the outward recession of optically thick gas at the edge of an expanding low-density inner cavity. In the very last epoch, the results are consistent with an increasing mass at larger radii, when the warm component increases in mass with time. The system may be getting ready again for another stronger release of gas towards the inner cavity. We expand on this idea in the next section.

\subsection{A longer-term modulation in gas accretion through the cavity?}

Models of eccentric binaries generally predict that variability in the total accretion rate onto the binary should be modulated by pulsed bursts on the timescale of one orbital period ($P_{bin}$) when the binary approaches periastron \citep{2016MunozR,miranda2017, 2015dunhill}. 
Strikingly, while atomic gas emission in the HI lines and the 5~\mic continuum follow this scenario, with enhancements at each periastron passage \citep[][and this work]{Tofflemire25}, the emission from all the hot molecular tracers observed with MIRI exhibits a longer-term modulation that does not simply follow the binary-induced pulsed bursts (see also Appendix \ref{app: additional_plots}). In Figure \ref{fig: disc+cartoon}, we show the longer-term modulation observed in this work in the broader context of the extended accretion time series that combines all the LCO photometry from this program \citep{Tofflemire25} and the publicly available datasets included in \citet{kospal25} that cover later accretion bursts observed in that work (Appendix \ref{app: accretion_history}). This combined dataset shows that the variability measured in the molecular lines reflects more the modulation of the burst peak intensity than the occurrence of each burst at periastron. To better highlight this potential global trend, we overplot a sinusoidal oscillation aligned with the modulation in the peak intensity of the bursts as well as the modulation in the molecular lines, which tentatively match a semi-stable period close to $\sim5$~$P_{bin}$.

A periodic modulation in the peak stellar accretion rate at each burst over $\sim5P_{bin}$ has been predicted in several simulations of circular binaries ($e=0$) \citep{2016MunozR,miranda2017,2012shi}, which commonly produce $m=1$ asymmetries in the surface density Fourier transform, designated as ``lumps", at the inner disk edge. The lumps, which are fed by the outer disk, eventually become unstable and decay onto the binary leading to accretion bursts with enhanced mass accretion over multiple binary periods, until the next ``lump" builds up. A potential periodicity of 4--5~$P_{bin}$ was tentatively identified in H-K band periodograms by \citet{Muzerolle19}, who suggested it may be connected to the ``lump" proposed by models. This longer-term modulation, if periodic, would correspond to a Keplerian radius of $\sim 0.4$~au, near where the inner disk edge is expected to be in dynamical models (Section \ref{sec: intro} and Table \ref{tab: DQTau props}). This region is also consistent to where we infer an increase in emitting gas mass with time in the warm water reservoir from the slab fits (Figure \ref{fig: Slab_results_+radii}), which could be interpreted as the gas mass building up again in a ``lump", getting ready to enhance the accretion rate in the next periastron passages. In support of this scenario, an increase in the peak accretion is in fact observed in the LCO light curve after our MIRI monitoring has ended (Figure \ref{fig: disc+cartoon}). 

A problem with this interpretation is that models do not predict the periodic lump formation and its modulation of the stellar accretion rate in eccentric binaries like DQ~Tau \citep{2016MunozR,miranda2017,2012shi}, except in one study by \cite{2011sytov} who found a $\sim$5~$P_{bin}$ modulation of the inner disk edge radius. Some models did find a modulation in the accretion peak intensity across multiple periastron passages \citep[see e.g. Figure 4 in][]{2016MunozR}, but did not find it to be periodic and did not investigate further their origin. This effect might in fact be consistent with what observed in this work, which is not strictly periodic (Figure \ref{fig: disc+cartoon}); this might prompt to revisit these effects in dynamical models of close binaries.

In comparing our results to these dynamical models, there are a few considerations to keep in mind. One is that the modulation we observe may be a transient, quasi-periodic modulation that emerges from the complex inner disk dynamics on the timescale of at most $\sim 10$ orbits or so, before being disrupted by binary interactions with the inner disk and perhaps re-emerge at later times. Historical records of photometry in DQ~Tau in fact have not recovered a stable 4--5~$P_{bin}$ periodicity on long timescales (Appendix \ref{app: accretion_history}), and our light curve in Figure \ref{fig: disc+cartoon} demonstrates that the modulation in burst peak intensity is certainly not simply periodic. However, we find it consistent with the later epochs observed in \citet{kospal25} several months after out monitoring, suggesting that this modulation may be relatively stable over yearly timescales. The evidence for a similar modulation from over a decade ago \citep[data from 2012--2014 in][]{Muzerolle19} suggest that the process may repeat across several years. Another consideration is that binary accretion models have typically focused on the gas accretion rate on the stars rather than the gas mass filling the region within the inner disk edge, which is instead what is observed with the hot molecular gas tracers studied in this work. 

To guide future comparisons to dynamical models, we summarize here what emerges from this monitoring as illustrated in Figure \ref{fig: disc+cartoon}. At the start of our MIRI monitoring campaign, immediately after periastron, the innermost disk region is filled with hot molecular gas down to the binary orbit. The molecular lines remain broad and their luminosity high in between and through the second burst, indicating that the innermost region is still filled with gas even after the first strong burst. This may explain the unusual early start and longer duration of the second strong burst (see Appendix \ref{app: accretion_history}, Figures \ref{fig: dqtau_LCO_archival} and \ref{fig: dqtau_LCO_archival_CDF}). Right before the third accretion burst, the stars are now at apastron and the accretion rate, the hot water luminosity, mass, and line widths are all at or close to their minimum, implying the innermost disk region is now mostly devoid of molecular gas out to $\sim 0.3$~au.

In the third accretion burst covered in our monitoring, the stars are at periastron but the burst has a much lower peak than the first two. The molecular gas tracers are still at their minimum luminosity, temperature, and mass, and their line widths show the gas is still at larger radii than during the first two bursts, with a temporary modulation shown by both the MIRI and iSHELL lines to broaden (i.e., material moving radially inward) at the peak of the third burst before narrowing again. This accretion burst may be showing us cases where the mass loaded into the accretion streams is close to a minimum and could explain the weak bursts that have been recorded in DQ~Tau (including right before the beginning of our MIRI monitoring and at orbital cycles 6--9 as labeled in Figure \ref{fig: disc+cartoon}); these weak bursts are known to happen $\sim 15\%$ of the time \citep[][and Appendix \ref{app: accretion_history}]{Tofflemire17}, consistent again to a periodicity of $\sim 5 P_{bin}$ if evenly distributed. 

\begin{figure*}
\centering
\includegraphics[width=0.8\textwidth]{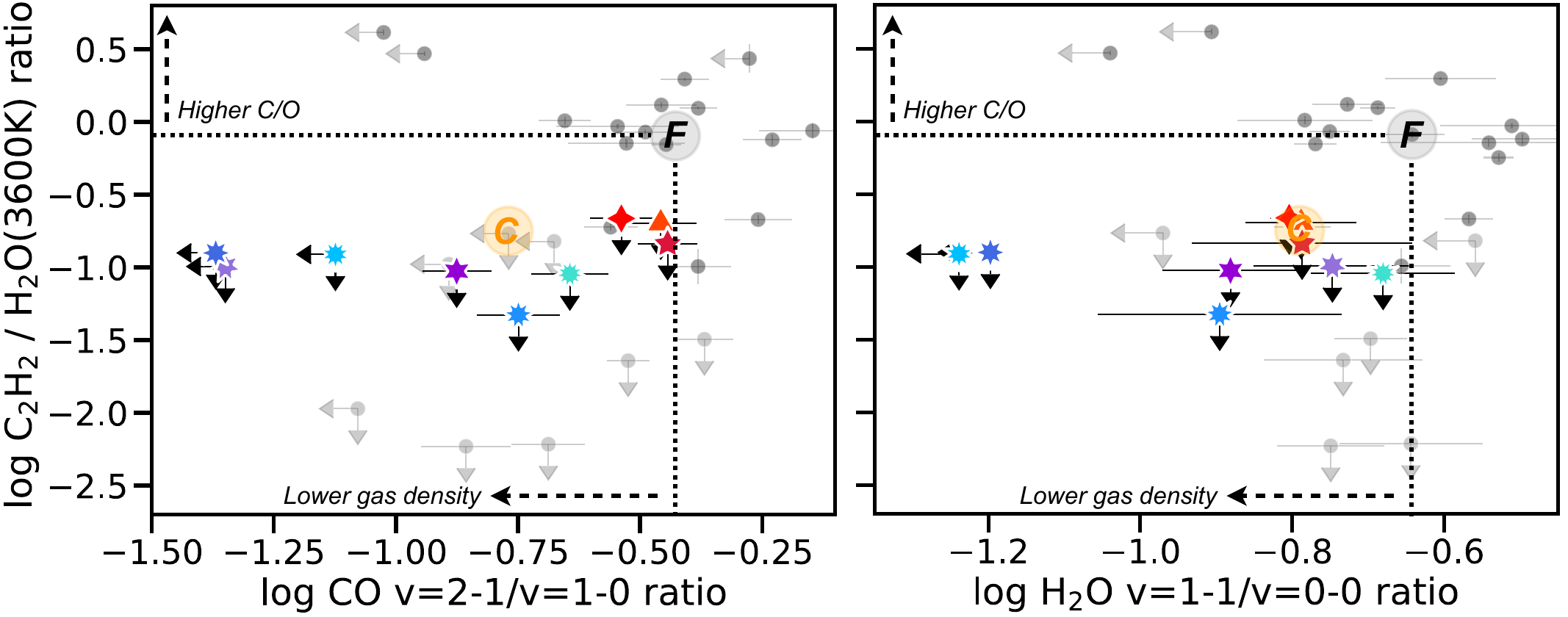}
\caption{Same as Figure 9 in \citet{mallaney26}, but showing DQ~Tau epochs in reference to the JDISCS-C1 sample of full disks (grey datapoints). The vibrational line ratios for CO (left) and \ce{H2O} (right) are sensitive to the gas density through the different critical density of higher-energy transitions. The dashed arrows near each axis indicate the direction of increasing the elements C/O ratio and decreasing the gas density in reference to the median value of full disks (larger grey circle marked with ``$F$"). The line ratios measured in DQ~Tau fall in the same region as values measured in disks with an inner dust cavity in \citet{mallaney26}, whose median is marked with the larger orange circle with ``$C$".}
\label{fig: CtO_density_diagram}
\end{figure*}

In the last epoch covered in our monitoring, the stars are at apastron again and yet we observe a sudden increase in the hot molecular gas luminosity. From the MIRI and iSHELL line widths, we can see that the hot molecular gas is moving in again, showing an increased hot and warm water mass (Figure \ref{fig: Slab_results_+radii}). A gas mass build-up at $\sim 0.3$~au in this phase may be connected to a longer-period accumulation of mass that prepares the system to accrete once again more strongly at the next periastron passage, as discussed above. 

An open question for future work is whether the molecular gas emission we observe with MIRI and iSHELL is tracing more the accretion streams or the global filling of gas in the inner cavity produced by the binary, or possibly a combination of the two regions, depending on the gas density and molecule survival to dissociation. Since the high-velocity CO gas extends to, or even into, the binary orbit at all times during our monitoring, the recession of CO and \ce{H2O} to larger radii could also be due to molecule dissociation in low-density gas, which would still require a modulation of the gas density. Even if these details have not been computed yet by models for direct comparison to the molecular spectra, it remains the case that the modulation of molecular gas luminosity and inner radius measured in this work follows a longer timescale than the individual binary period, in contrast to what models predict for the stellar accretion rate.

\subsection{Implications for \ce{C2H2} formation and survival}
A time-dependent filling of an inner disk cavity in DQ~Tau may have implications for the formation and survival of organic molecules. Recent work by \citet{mallaney26} found that disks with an inner dust cavity on average have lower CO $v$=2-1/$v$=1-0 and \ce{H2O} $v$=1-1/$v$=0-0 line ratios as measured in MIRI spectra in comparison to those observed in full disks without a cavity. These ratios are sensitive to the volume gas density: when the gas density is much lower than the critical density necessary to thermalize the emission by collisions, the higher-vibrational lines (which have higher critical densities) are sub-thermally excited and these ratios decrease. Figure \ref{fig: CtO_density_diagram} shows the diagrams of these line ratios and the \ce{C2H2}/\ce{H2O} ratio, which is instead a proxy for the elemental C/O ratio in the inner disk \citep{tabone23,Colmenares24,Arabhavi25,Arabhavi26,grant25,long25}. 

The CO line ratio measured in DQ~Tau as a function of time is at first consistent with the median of full disks in the first three epochs, when the cavity is filled with molecular gas (see Figure \ref{fig: Kepl radii}), suggesting a higher molecular gas density at $\sim$~0.1--0.2~au.
The following five epochs show decreased CO $v$=2-1/$v$=1-0 line ratios, populating the region of disk cavities in \citet{mallaney26}. At the same time, narrower line widths suggest the emission is moving out to larger radii. This can be understood as a decrease in the gas density in the inner $\sim$~0.1--0.2~au, in connection to the decrease in the high-energy line luminosity and gas mass of the hot component. In the last epoch the CO ratio increases again towards the median of full disks. At the same time, the water high-energy line luminosity in DQ~Tau suddenly increases again and the line widths of the intermediate-energy lines broaden indicating they move back in (Figure \ref{fig: line_var}) and that the emitting gas mass increases (Figure \ref{fig: Slab_results_+radii}). All these changes in the last epoch are consistent with a gas density that increases again in the inner cavity.

The \ce{H2O} line ratio is more uncertain but clusters around or below the median value measured in cavity disks in \citet{mallaney26} in all DQ~Tau epochs.
As discussed in modeling papers before, a lower gas density and a decreased UV shielding from water would all conspire against the presence of organics and in particular \ce{C2H2} in the upper disk layers, by decreasing the formation rates and increasing UV photodissociation \citep{duval2022,Woitke2024,Kanwar2024}. This scenario could explain the only common molecule that is missing in DQ~Tau in comparison to other T~Tauri disks around single stars of similar age and mass: \ce{C2H2}. 

\begin{figure}
\centering
\includegraphics[width=0.45\textwidth]{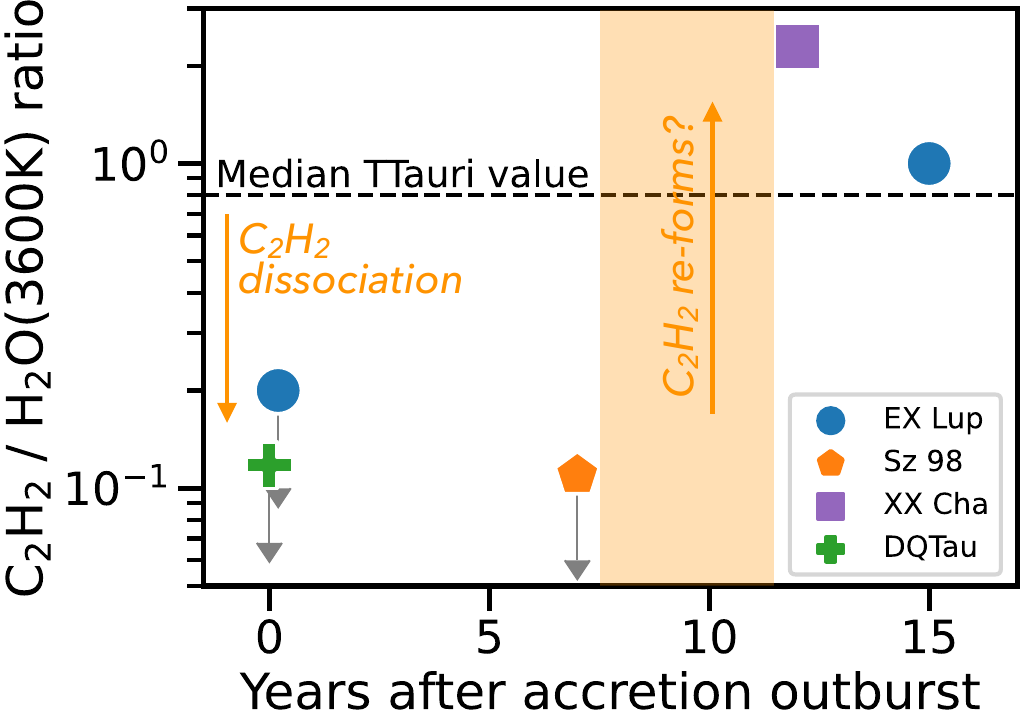}
\caption{Potential connection between \ce{C2H2} and accretion outbursts: \ce{C2H2} could be UV photodissociated during outbursts and take $\sim 10$~yr to re-form in the inner disk. The value shown for DQ~Tau is the average of all epochs included in this work. The median T~Tauri value shown in black is taken from Figure \ref{fig: CtO_density_diagram}.}
\label{fig: C2H2_reform}
\end{figure}

Another related aspect that it is interesting to consider is illustrated in Figure \ref{fig: C2H2_reform}. In addition to a lower gas density and a decrease in UV shielding by water as discussed above, episodes of enhanced UV irradiation may also play a role in destroying \ce{C2H2} in inner disks. In Figure \ref{fig: C2H2_reform} we present three systems where strong accretion outbursts have been recorded in the past 15 years and that have MIRI spectra available. EX~Lup has been analyzed in \citet{2025smith}, Sz~98 in \citet{gasman23}, and XX~Cha in \citet{grant25}, with the recent accretion outburst measured in \citet{claes22}. We have processed their MIRI spectra with the same pipeline and analysis as the DQ~Tau epochs in this work to be fully consistent in the comparison. These systems show that the \ce{C2H2}/\ce{H2O} flux ratio is low within $\sim 7$ years from the most recent outburst, where we include also DQ~Tau due to its continuous accretion bursts. After 10 years, two systems show that \ce{C2H2} is detected again and the \ce{C2H2}/\ce{H2O} ratio is consistent with values measured in T~Tauri disks that do not have strong accretion outbursts. These two systems include one case where the spectra were measured both during and after a strong outburst, EX~Lup, which so far is the only case showing evidence for the potential re-formation of \ce{C2H2} in the same disk. This small sample of disks suggests an interesting direction for future monitoring and modeling studies, to determine the impact of accretion bursts and the re-formation timescales of \ce{C2H2} in inner disks.

\subsection{Accretion-induced ice sublimation events} \label{sec: disc_ice_subl}
One of the fundamental questions that this monitoring program had at its conception was to study the effects of strong accretion bursts in causing episodes of enhanced ice sublimation at the snowline, and their potential consequences on planet-forming solids \citep[e.g.][]{houge2023,ros2024}. This was inspired specifically by previous work on the Class II disk in EX~Lup, where enhanced cold water emission near the snowline was observed during and after a strong accretion outburst \citep{banz12,2025smith}, suggesting that accretion-induced cold water layers may last for up to a decade after an outburst \citep[see also modeling and discussion in][and in Krijt et al. in prep.]{Vlasblom25}. In DQ~Tau, there is evidence for a fluctuation in the 1500~K line luminosity at 20--40 days after the first strong burst (Figure \ref{fig: line_var}) as well as for the detection of $\sim 200$~K water emission and its average increase later in the monitoring (Figure \ref{fig: diagnostic_diagrams}). This trend is in the direction expected from a fluctuation of the snowline to larger radii as pushed by an accretion outburst \citep[Figure 8 in][]{2025smith}, but it is much smaller than observed in EX~Lup \citep[Figure 5 in][]{2025smith}, despite the accretion luminosity increase being similar in the first accretion burst observed in DQ~Tau in this campaign. The different duration of the accretion events in the two systems, which lasted several months in EX~Lup in comparison to a maximum of $\sim6$ days in DQ~Tau, is likely the origin of the different chemical reaction in the two disks. 

\begin{figure}
\centering
\includegraphics[width=0.45\textwidth]{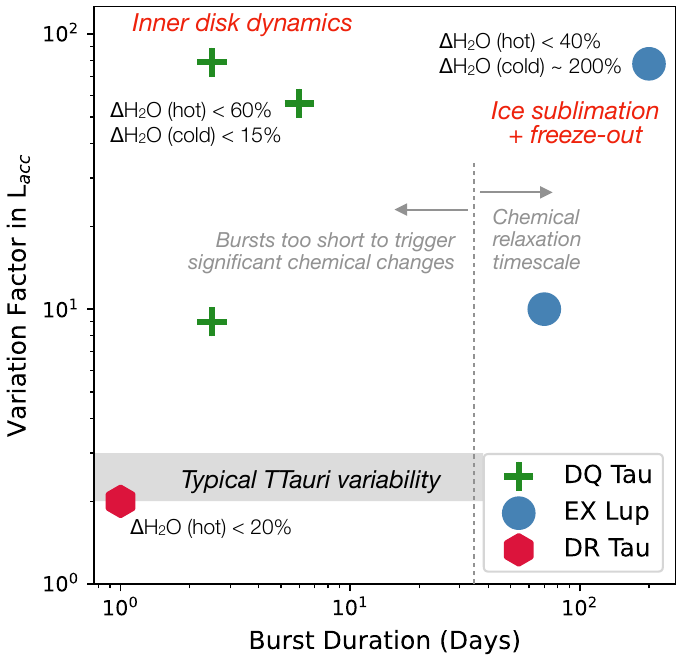}
\caption{Variation in accretion luminosity as a function of burst duration. Cold water emission from a reservoir near the snowline was observed to strongly increase in EX~Lup during a several-months-long outburst in 2008 \citep{banz12,2025smith}. DR Tau was found to exhibit no change in hot-water emission with a factor $\sim 2$ change in accretion \citep{banz14}. This new DQ~Tau monitoring fills in a unique region in this parameter space with strong but short accretion outbursts, which are found to modulate the hot water reservoir with very limited influence on the cold reservoir. The values in the figure report the maximum variation measured in hot and cold water in each case (the global change across three bursts in the case of DQ~Tau); the chemical relaxation timescale from \citet{Woitke2009,Woitke2024} is shown for reference.}
\label{fig: Outburst_amplitude_duration}
\end{figure}

This is summarized in Figure \ref{fig: Outburst_amplitude_duration}, where we combine the three bursts observed in DQ~Tau in this monitoring \citep{Tofflemire25} to two recent ones measured in EX~Lup in 2008 and 2022 \citep{cruzsaenzdemiera23,wang23} and another similar experiment done on DR~Tau, which was observed to monitor the accretion variability more typical to T~Tauri stars \citep{banz14}. This earlier experiment found that in the case of accretion variability confined within a factor of 2, the mid-infrared water lines did not change by more than 20\%. However, that work only observed a few high-energy lines ($E_u$~5000--6000~K) that are not sensitive to emission from the snowline. In EX~Lup, the 2008 outburst lasted almost 1 year before falling into a quiescent state for more than a decade and was found to produce a ``burst" in cold water emission from near the snowline, consistent with it being shifted to larger disk radii during the outburst \citep{banz12,2025smith}. In DQ~Tau, the accretion bursts have similar amplitude but are both much shorter in duration, less than one week, and more frequent (every 16 days). The lack of similarly strong cold emission and variability in DQ~Tau in comparison to EX~Lup therefore is likely attributed to the disk not having enough time to reach equilibrium with the increased irradiation during outburst (see Section \ref{sec: disc_singlestarstrends}). As found and discussed in this work, the molecular variability in DQ~Tau seems to be dominated by dynamical effects more than excitation or chemical effects.

\subsection{Extending the analysis to four additional epochs from \citet{kospal25}}
Lastly, we discuss our results in the context of another recent monitoring of DQ~Tau published in \citet{kospal25}, which we refer to with K25 in this section for brevity. K25 obtained 4 epochs of MIRI spectra between January 29 and March 2 (a few months after the end of our MIRI monitoring), spanning three orbital periods similar to our monitoring and sampling 2 burst peaks and 2 low-accretion epochs in between them (Figure \ref{fig: disc+cartoon} and Figure 3 in K25). We have re-reduced and analyzed the MIRI spectra included in K25 with the same tools and procedures as our data to ensure the most reliable comparison. The measured properties of the K25 epochs overlap very well with our first 3 epochs of data in this work, both in terms of line luminosity, radii, and slab-fit component results (Figures \ref{fig: Slab_results_+radii} and \ref{fig: disc+cartoon}). This suggests that the K25 dataset sampled DQ~Tau under conditions very similar to the beginning of our monitoring, when the inner disk cavity was filled with hot molecular gas. In comparison to the analysis and proposed interpretation in K25, by combining all the MIRI epochs in both works (a total of 13) it is now possible to clarify some results, as follows.

While K25 reported a tentative correlation of molecular line excitation with \lacc based on 4 epochs, the combined 13 epochs show no correlation with \lacc. As discussed above, the molecular gas luminosity and temperature reflect the change in orbital radius and area of the emission that vary over a timescale longer than the individual orbital period, rather than responding directly to the change in \lacc. The combined 13 epochs are consistent with the possibly quasi-periodic modulation over 4--5~$P_{bin}$ discussed above in this section, previously proposed in \citet{Muzerolle19} and shown in Figure \ref{fig: disc+cartoon}. The existence of this longer-term modulation tentatively reconciles all observations over several years, including those from \citet{Muzerolle19} and the two MIRI campaigns presented in K25 and in this work.
        
The modulation in temperature, luminosity, and radius of an inner disk dust rim proposed in K25 is consistent with the modulation in molecular gas emission found in this work; to support that, we note that the Keplerian radius we estimate from water lines in the K25 epochs follows the same modulation they report for the inner dust radius inferred from fits to the dust temperature, where the radius moves out in the second epoch, moves in in the third, and moves out again in the fourth, suggesting that dust and gas move in and out together. We should note that binary models do not expect a stable inner rim at a specific radius \citep[e.g.][]{2016MunozR} and both gas and dust observations should in the future be more directly interpreted in the context of asymmetric inner disk structures and accretion streamers, which are not accounted for in our analysis nor in K25.
    
The variable shadowing of the outer disk proposed in K25 to explain the variations in dust emission possibly explains the trend found in the 13~$\mu$m continuum in this work (opposite to the trend in the hot molecular gas), where the continuum flux increases with time with a maximum when the hot gas luminosity is instead minimum and has receded out to $\sim0.3$~au. A recession of hot gas and dust may allow a higher irradiation of colder dust from larger radii, in a similar scenario as proposed in \citet{2025Chengyan}.
    
Rather than a C-poor disk as proposed in K25, by comparison to \citet{mallaney26} our analysis suggests a lower inner gas density to explain the absence of \ce{C2H2} in the inner disk of DQ~Tau; this is supported by recent modeling that identifies the relative strength of \ce{CO2}, which is higher in DQ~Tau (Figure \ref{fig: dqtau_correlations}), as a C/H tracer \citep{Arabhavi26}, suggesting its disk may actually be C-rich. Another interpretation for the stronger \ce{CO2} emission may be the small inner disk cavity, as proposed in \citet{Vlasblom24}.

\section{Summary \& Conclusions} \label{sec: conclusion}
In this monitoring program, we have obtained and analyzed 9 epochs of JWST-MIRI spectra and 9 epochs of IRTF-iSHELL spectra of the binary system DQ~Tau across 3 sequential periastron-induced accretion bursts (Section \ref{sec: intro}). For completeness and comparison, we also analyzed 4 additional MIRI epochs collected a few months after the end of our monitoring and recently published in \citet{kospal25}.
The main results and conclusions from this work are as follows:

\begin{itemize}
    \item \textit{Different modulations in the inner disk of DQ~Tau}: While the variability in atomic gas (infrared HI lines) and hotter inner disk dust (5~$\mu$m continuum) closely follows the accretion rate as regulated by the binary orbit, all the disk molecular gas tracers show longer-term trends in both luminosity and radial emission with only a minor modulation at individual periastron passages. Specifically, the hot gas tracers (including CO, H$_2$O, HCN, OH) globally weaken across multiple orbital periods, while the colder dust and gas tracers show the opposite luminosity trend.
    
    \item \textit{Dynamical effects on molecular emission}: Time-resolved Doppler mapping of water lines measured via line broadening from MIRI spectra shows that their emission is moving to larger radii over time, if we interpret the line widths as dominated by Keplerian rotation in the circumbinary disk. This is confirmed by the fully-velocity-resolved CO line profiles measured from iSHELL spectra (Figures \ref{fig: main_display} and \ref{fig: Kepl radii vs time}). This radial evolution can explain the decrease in line luminosity, temperature, and emitting area for the hot molecular gas in a unified picture.

    \item \textit{Long chemical timescales in the molecular region}: The variation in molecular line luminosity measured in DQ~Tau does not follow the global correlations with \lacc defined by a larger sample of disks around single stars (Figure \ref{fig: dqtau_correlations}), nor the large increase in cold water emission observed in EX~Lup during a similar-intensity accretion burst (Figure \ref{fig: Outburst_amplitude_duration}). This demonstrates that the shorter-term bursts in DQ~Tau do to not trigger significant chemical changes in the inner disk despite their strength. This result supports the long chemical relaxation timescale that is expected by models, with implications for inner disks around single stars too: the hot molecular emission may depend on average irradiation conditions rather than respond immediately to their short-term (up to $\sim 10$~days) accretion variability. 

    \item \textit{Organic formation and survival}: The relatively low CO $v$=2-1/$v$=1-0 and H$_2$O $v$=1-1/$v$=0-0 line ratios in DQ~Tau are similar to those found in disks with a dust cavity in \citet{mallaney26} and similarly suggest a lower gas density and a decreased UV shielding that can cause a decrease in organic abundance in the inner disk, in particular for C$_2$H$_2$ (Figure \ref{fig: CtO_density_diagram}). We propose this scenario as the origin of the weak C$_2$H$_2$ emission in DQ~Tau.

    \item \textit{Longer-term modulation of molecular tracers}: We propose that a longer-term quasi-periodic modulation of $\sim 5 P_{bin}$ is consistent with the inner disk molecular gas luminosity and burst peak accretion modulation (Figure \ref{fig: disc+cartoon}). This is reminiscent of the lump-modulated accretion strength in dynamical simulations of low-eccentricity binaries; a similar modulation of $\sim 5 P_{bin}$ has been previously reported based on $H-K$ variability in \citet{Muzerolle19} and is consistent with the later 4 epochs from \citet{kospal25}.
    
\end{itemize}

The results from this monitoring program support previous findings that inner disk dust and gas tracers in DQ~Tau show a complex longer-term variability across multiple orbital periods that does not seem to be fully captured by existing dynamical models, which instead predict a modulation dominated by the single orbit with accretion pulses at each periastron passage. However, previous models did not directly provide the distribution of molecular gas within the inner region, which is what should be compared to the observations in this work. This motivates future modeling and monitoring efforts to better clarify the formation, excitation, and survival of inner disk molecules in the planet-forming region not only in disks around binaries but also around single stars, where the formation of a small inner disk cavity may produce similar conditions to those found in DQ~Tau.

\vspace{5mm}
\facilities{JWST, IRTF}

\software{
Matplotlib \citep{matplotlib}, NumPy \citep{numpy}, SciPy \citep{scipy}, Seaborn \citep{seaborn}, Astropy \citep{astropy:2013, astropy:2018, astropy:2022}, LMFIT \citep{lmfit}, iSLAT \citep{iSLAT,iSLAT_code,iSLAT_codev5}
}

\acknowledgments
The authors are grateful to the whole team that supports JWST operations and in particular to B. Perriello, B. Sargent, and T. Roman for their help during this JWST monitoring program to maximize the quality of the data and re-schedule two epochs missed by the observatory due to technical issues. 
The authors are grateful to Dr. Diego Mu\~{n}oz for several helpful conversations regarding the dynamical modeling of DQ~Tau, and to the anonymous referee for thorough comments that helped improve this work.
The authors acknowledge support from NASA/Space Telescope Science Institute grant JWST-GO-04727.
This work is based on observations made with the NASA/ ESA/CSA James Webb Space Telescope. The JWST data used in this paper were obtained from the Mikulski Archive for Space Telescopes (MAST) at the Space Telescope Science Institute and can be accessed via:
\dataset[10.17909/n9jx-1n57]{http://dx.doi.org/10.17909/n9jx-1n57}.
The data were obtained from the Mikulski Archive for Space Telescopes at the Space Telescope Science Institute, which is operated by the Association of Universities for Research in Astronomy, Inc., under NASA contract NAS 5-03127 for JWST. The observations are associated with JWST GO Cycle 3 program 4727.
Part of this research was carried out at the Jet Propulsion Laboratory, California Institute of Technology, under a contract with the National Aeronautics and Space Administration (80NM0018D0004). 
Sebastiaan Krijt and Till Kaeufer acknowledge support from STFC Grant ST/Y002415/1.
CFM is funded by the European Union (ERC, WANDA, 101039452). Views and opinions expressed are however those of the author(s) only and do not necessarily reflect those of the European Union or the European Research Council Executive Agency. Neither the European Union nor the granting authority can be held responsible for them.
CHR acknowledges the support of the Deutsche Forschungsgemeinschaft (DFG, German Research Foundation) Research Unit "Transition discs" - 325594231. CHR is grateful for support from the Max Planck Society.
We gratefully acknowledge the contributions of the AAVSO observer community, whose photometric data and metadata resources were used in this study and made available through the AAVSO’s scientific archives. We thank the following AAVSO Astronomers for contributing DQ Tau photometry: Enrique T. Boeneker, Matthias Kolb, Gordon Myers, Kenneth Menzies, Robert Stephens, Bill Tschumy, Ray Tomlin, and Mike Wiles.

%%%%%%%%%%%%%%%%%%%%%%%%
\appendix

\section{Accretion Burst Statistics from Archival Photometry} \label{app: accretion_history}

To place the accretion bursts we observe during our \jwst\ monitoring campaign in the broader context of DQ Tau's variability, we compile photometry spanning more than 10 years to construct an archival accretion-luminosity time series. To supplement the LCO $u^\prime$ and X-shooter measurements described in this work, we include the following data sets: LCO $U$-band photometry from \citet{Tofflemire17} and from observations in the 2023-2024 observing season, $g$- and $V$-band photometry from the ASAS-SN project \citep{Shappee2014}, $B$- and $V$-band photometry from AAVSO during the 2024-2025 season, and $B$-band photometry from the UVOT instrument on Neil Gehrels Swift Observatory \citep{Gehrels2004}. The latter two provide high-cadence data sets covering the accretion bursts observed in the \citet{kospal25} JWST program. 

ASAS-SN photometry was re-extracted from the Sky Patrol database, which is documented in \citet{Shappee2014,Kochanek2017}. AAVSO photometry was retrieved from the AAVSO International Database (AID). Only measurements transformed into the standardized AAVSO magnitude system (e.g., corrected for airmass, etc) are considered. Both ASAS-SN and AAVSO data with nightly visit in short succession are binned into a single epoch. Reduced Swift UVOT $B$-band images were downloaded from the public archive. Differential ensemble photometry was extracted using the approach described in \citet{Tofflemire17,Tofflemire25}.

\begin{figure}
\centering
\includegraphics[width=0.5\textwidth]{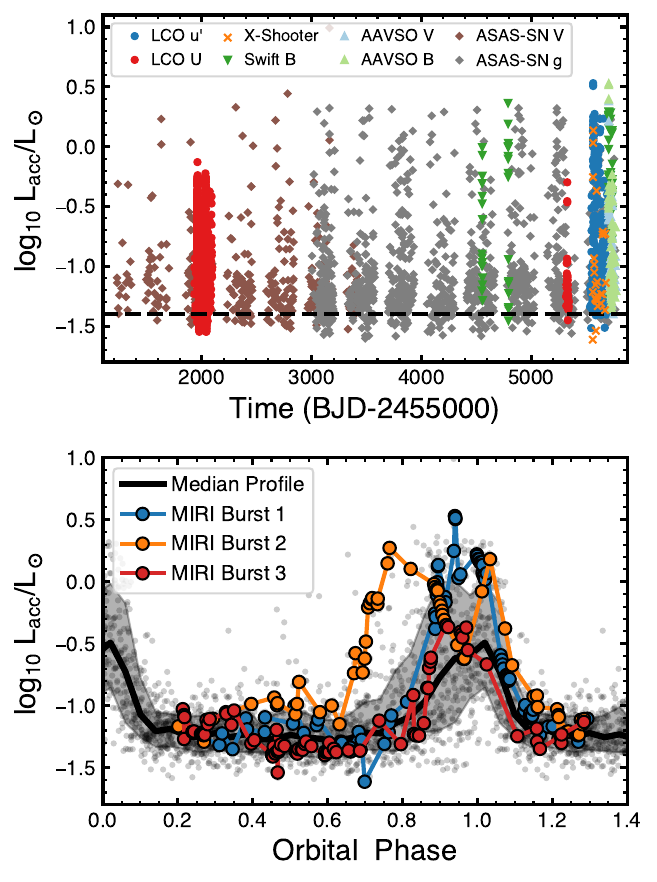}
\caption{DQ Tau archival accretion luminosity time series. {\bf Top:} Full accretion luminosity time series derived from X-shooter spectra, LCO $u'$ and $U$ photometry, ASAS-SN $g$ and $V$ photometry, AAVSO $B$ and $V$ photometry, and Swift $B$ photometry. {\bf Bottom:} Accretion luminosity phase-folded on the DQ Tau binary orbital phase. The time series has been binned to 8 hr. The median profile with its 68\% confidence interval are shown with the black line and gray shaded region, respectively. The three bursts covered during our \jwst\ campaign are over-plotted with colored lines, demonstrating the above-average strength of the first two bursts. }
\label{fig: dqtau_LCO_archival}
\end{figure}

To convert photometric magnitudes to an accretion-luminosity time series, we combine two approaches: one for the redder filters ($g$, $B$, $V$), which have significant temporal overlap with the \lacc\ measurements from our program (derived from X-shooter spectra and $u^\prime$ photometry) or with each other, and another for the $U$-band photometry, which does not have significant overlap with the other data sets. For the redder filters, we first map \lacc\ measurements from this work onto the ASAS-SN $g$ data. The same procedure is then applied mapping the ASAS-SN $g$ accretion luminosities onto the ASAS-SN $V$, AAVSO $B$, AAVSO $V$, and UVOT $B$ data sets. The mapping takes place in logarithmic space between the accretion luminosity and relative flux of the data sets in question. A polynomial of first or second order is used, informed by the Bayesian Information Criterion (BIC). Measurements must be separated by $<4$ h to be considered in the fit. The root-mean-square (RMS) of the residuals of the $u^\prime$ to ASAS-SN $g$ fit is 0.15 dex. The RMS of the ASAS-SN $g$ to ASAS-SN $V$, AAVSO $B$, AAVSO $V$, and Swift $B$ fits are 0.15, 0.07, 0.07, and 0.07 dex, respectively. This scatter contains both systematic uncertainties in the conversion and intrinsic variability in the accretion rate over the (up to) 4 h window; it is much smaller than the $>1.5$ dex of variability observed. 

The $U$-band data fall at times where overlap with the ASAS-SN data sets are sparse, making the mapping approach above poorly constrained. We convert these data to accretion luminosities using a zero-point offset and a scaling relation informed by our $u^\prime$ and X-shooter data sets. We set the U-band flux zero-point as the median value during binary orbital phases between 0.4 and 0.65 where accretion is consistently at its lowest (and likely near the detection floor; \citealt{Tofflemire25}) and assign it to the \lacc\ value from the same phase range in the $u^\prime$-derived \lacc\ time series ($-1.3$ dex). This choice is justified by the nearly constant baseline seen during these phases in the ASAS-SN light curves spanning $\sim$10-year timescales. The scaling relation is derived from X-shooter flux-calibrated spectroscopy using measured accretion luminosities and synthetic $U$-band photometry for each of the 25 DQ Tau epochs. From the limited number points that overlap with ASAS-SN accretion luminosities, we find a median offset (0.03 dex) that is less than and a residual RMS (0.15 dex). While these overlapping data do not span the full range of variability (only $\sim 0.9$ dex), they suggest the two approaches are self-consistent to a degree that is acceptable for this use case. The full accretion luminosity time series is presented in the top panel of Figure \ref{fig: dqtau_LCO_archival}. 

A few caveats in this exercise are worthy of note. First, we assume constant extinction. Although this assumption is well motivated from the X-shooter spectral fits \citep[measuring $A_V$ and \lacc\ simultaneously;][Table 1]{Tofflemire25}, we also find evidence for sporadic reddening events that are not accounted for here. Second, our goal is to achieve a self-consistent \lacc\ time series to investigate relative changes in DQ Tau's accretion variability. For this reason, we derive our own \lacc\ values for the ASAS-SN, Swift, and AAVSO data sets rather than adopting values from \citet{kospal25}. The basis for our \lacc\ system is the X-shooter spectra, which have an estimated absolute error of $\sim0.2$ dex \citep{Manara2017}. This analysis will be developed in more detail with additional data sets in an upcoming publication (Tofflemire et al., in prep). 

\begin{figure}
\centering
\includegraphics[width=0.5\textwidth]{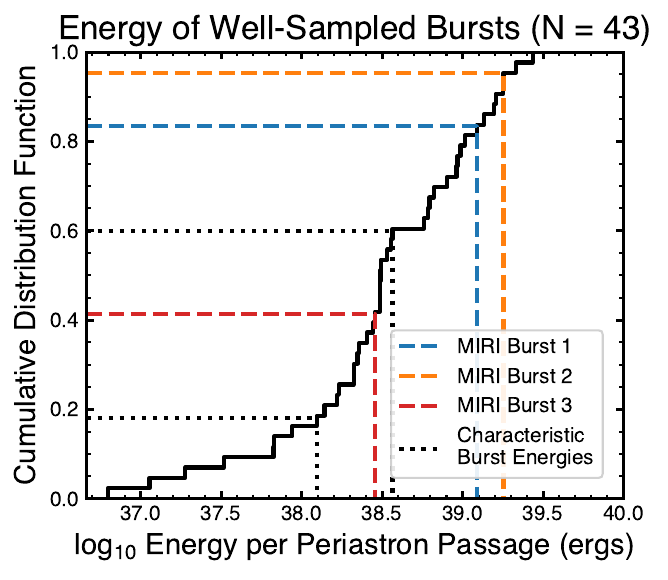}
\caption{Cumulative distribution of integrated burst energies from 43 well-sampled periastron passages. Dotted lines mark the range of typical burst energies as defined by the steep rise in the CDF. Of the three bursts observed in our \jwst\ monitoring program, which are marked with dashed colored lines, the third falls into the typical energy range while the first two fall well above it, in the range of strong outbursts (see also Figure \ref{fig: dqtau_LCO_archival}). Low-amplitude bursts occur $\sim$15-20\% of the time. }
\label{fig: dqtau_LCO_archival_CDF}
\end{figure}

In the bottom panel of Figure \ref{fig: dqtau_LCO_archival} we present the phase-folded data set described above and the median burst profile with the 68\% interval shown in the shaded region. In this plot, and in the computation of the median profile, we have binned the input time-series to 8 hours as to not bias the profile to times of dense monitoring campaigns that aimed to detect short-timescale variability. By comparing to the \lacc\ time series of the bursts monitored by our \jwst\ campaign, this analysis demonstrates that the first two bursts were in the high-energy end of what is seen in historical records, while the third was consistent with the median burst amplitude and duration. An additional peculiarity of the second burst was its earlier than typical start, at about an orbital phase of 0.6 when bursts are rare in historical records. This is consistent with the general picture emerging from this work as presented in the main text above, where the innermost disk kept being filled with gas between the first two accretion bursts (Figure \ref{fig: disc+cartoon}) probably causing the early start, high intensity, and longer duration of the second burst.

For a subsample of well-sampled orbital periods, we integrate the accretion luminosity to compute the total energy emitted in each accretion burst. Here we consider orbits where at least 8 measurements are present between orbital phases 0.65 and 1.2 (8.7 days), which encompasses most burst durations. There are 43 well-sampled bursts that meet this criterion in our sample. A cumulative distribution function (CDF) of the burst energies is presented in Figure \ref{fig: dqtau_LCO_archival_CDF}. The CDF reveals a range of characteristic burst energies (dotted lines), demonstrated by the steep rise in the CDF function, and a set of high- and low-amplitude bursts. The colored dashed lines mark the three bursts covered in our \jwst\ program, again demonstrating the above-average strength of the first two bursts and the typical strength of the third one, consistent with what is shown in Figure \ref{fig: dqtau_LCO_archival}. The CDF analysis also demonstrates that low-amplitude bursts occur $\sim$15-20\% of the time, or once every 5-6 orbits, on average. This is consistent with the longer-term modulation highlighted in Figure \ref{fig: disc+cartoon} and provides circumstantial evidence for a dynamical process that regulates the flooding of the inner disk cavity on a timescale longer than the individual orbital period, as discussed in Section \ref{sec: discussion} for the interpretation of the main results of this work.

\section{Updated HI-\lacc relations} \label{app: updated_Lacc_relations}
In this section, we provide the updated relations for the conversion of HI line luminosities measured from MIRI spectra into estimates of \lacc, previously presented as part of this observing program in \citet{Tofflemire25}. The reason for this update is that the new JDISCS reduction v9 significantly improved the absolute flux calibration across MIRI spectra, in turn changing the measured line fluxes within approximately -5\% (at shorter wavelengths) and 15\% (at longer wavelengths), as already reported in \citet{mallaney26}. In addition to the updated data reduction, we fit the relations using the combined sample of DQ~Tau and the larger T~Tauri disk sample from the JDISCS \citep{Arulanantham25} and MINDS \citep{MINDS24} collaborations as previously included in \citet{Tofflemire25}, here updated to the JDISCS v9 reduction and homogeneously analyzed as described above for DQ~Tau.

\begin{figure*}
\centering
\includegraphics[width=0.8\textwidth]{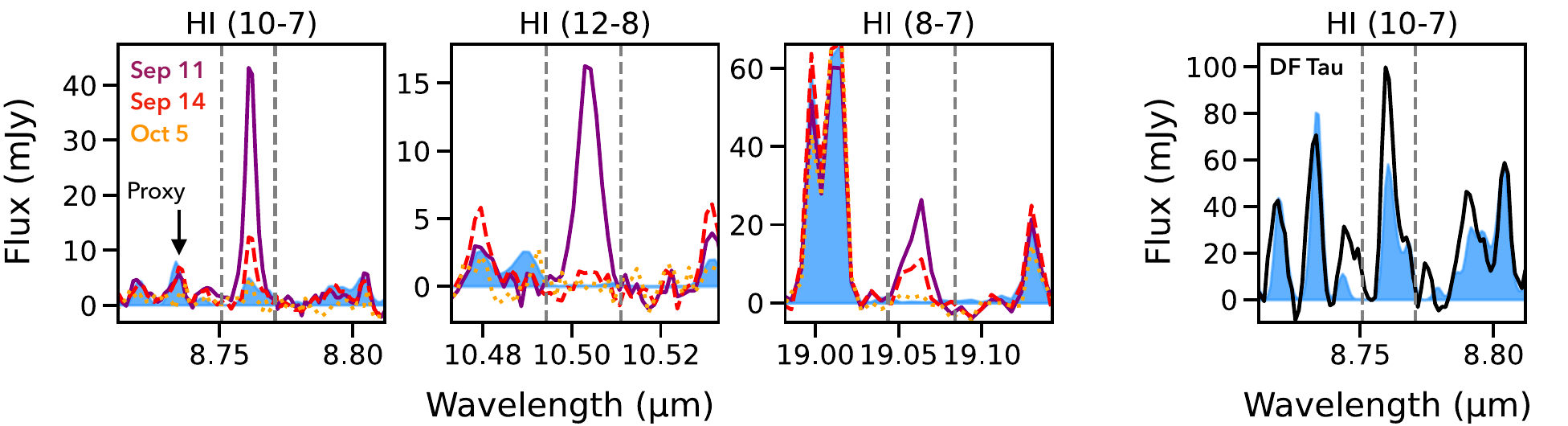}
\caption{HI lines used for the updated accretion relations shown in Figure \ref{fig: AccRelations_updated}, using three epochs of our DQ~Tau monitoring to illustrate the full range of accretion levels covered in this work. A water slab model in LTE is shown in light blue in the background for reference to the contaminating lines. While the (12-8) and (8-7) lines are completely uncontaminated, the (10-7) is often contaminated at the 10\% level and in some cases up to $\sim 70\%$ of its flux, depending on the strength of accretion and of the ro-vibrational emission lines. To the right, we report the extreme case of DF~Tau \citep{grant24}, where the estimated contamination of the (10-7) line is at the $\sim 70\%$ level. To correct for water contamination, we use the nearby emission line blend at 8.73~\mic (marked with an arrow) as an empirical proxy for the strength of the emission line blend under the (10-7) line.}
\label{fig: HI_lines_updated}
\end{figure*}

In comparison to \citet{Tofflemire25} we make two updates to our analysis of HI lines. First, we drop the (7-6) line, which is generally strongly contaminated by water emission \citep[][estimates a median contamination of 50\%, ranging between 30\% and 90\%]{Shridharan25}. Even if a line-proxy correction was applied in \citet{Tofflemire25}, the strong contamination makes the (7-6) line too unreliable for general use in lower-accretion sources that still have significant water emission, also due to the very limited pixel sampling of MIRI spectra that make the water subtraction process very dependent on the details of the analysis (water slab fits, RV shifts, interpolation and subtraction). We instead try to rely on the least contaminated lines, which have already been identified in \citet{banz25}: the (12-8) and (8-7) lines that are completely free from any contamination from molecular emission (to the best of our knowledge), and the (10-7) line that was identified as having minimal contamination (Figure \ref{fig: HI_lines_updated}). We remark that we do not attempt to use any other HI lines due to their higher contamination from water and/or OH, including the (6-5), (8-6), (9-6), and (13-8) lines that have been used in \citet{Shridharan25}. We consider these lines too unreliable and prone to provide spurious accretion values, especially in lower accretors.

After more extensive inspection of the MIRI spectra observed in the larger disk sample across different accretion levels we find that even the (10-7) line can be significantly contaminated by water emission, typically at the $\lesssim 15\%$ level but up to the 70\% level in the worst cases (still generally lower than the contamination of the (7-6) line). During our DQ~Tau monitoring, we estimate a contamination of the (10-7) line, when detected, between a minimum of 6\% (on October 11) and a maximum of 70\% (on September 14). To correct for water contamination, we subtract the flux of a nearby emission line blend at 8.73~\mic, selected as an empirical proxy for the strength of the emission line blend under the (10-7) line based on their similar total flux (Figure \ref{fig: HI_lines_updated}). This is still an approximation but it is the best that can be done without a detailed fit to the water emission. The water transitions that emit significantly at these wavelengths are ro-vibrational transitions from the $v=1-0$, $v=2-1$, or $v=3-2$ bands, which are hard to characterize due to the different non-LTE excitation of different bands as already reported in \citet{banz25}.  

\begin{figure*}
\centering
\includegraphics[width=\textwidth]{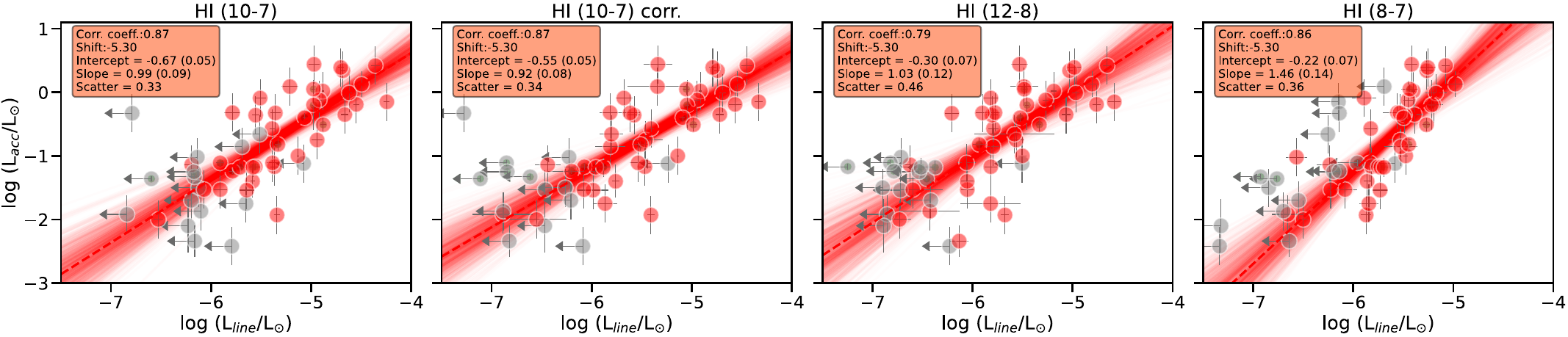}
\caption{Accretion luminosity relations using MIRI HI lines, updated to JDISCS reduction v9. Red datapoints are detections in the combined sample of DQ~Tau epochs from this work and T~Tauri disks observed in Cycle 1, grey are upper limits. Linear regression results are summarized in Table \ref{tab: lin_fit_params}. }
\label{fig: AccRelations_updated}
\end{figure*}

In Figure \ref{fig: AccRelations_updated} we report the updated accretion relations using the HI lines discussed in the previous paragraph. We include for reference the (10-7) line before and after correction for water contamination. The slope of the (10-7) line relation decreases from 0.99 to 0.92 after correction, but is still within 1~$\sigma$ of the confidence limits in the estimated regression parameters. It is also consistent within 1~$\sigma$ with the results reported in \citet{Shridharan25}, where water contamination was ignored. The slope of the (12-8) line is consistent with that of the (10-7) line within 1~$\sigma$, while that of the (8-7) line continues to be significantly larger as found in \citet{Tofflemire25}, even if lower than in that work.  We report all the updated relations parameters in Table \ref{tab: lin_fit_params}. 

We caution that estimating accretion from these lines as measured in MIRI spectra presents multiple challenges that have at least partially discussed in previous works \citep{Tofflemire25,Shridharan25} and in this section. Additional challenges that should not be ignored are the quality of continuum calibration, the possible presence of residual fringes at 8--10~\mic depending on the quality of each spectrum and of the fringe correction process \citep{pontoppidan24}, and photospheric absorption in low-veiling spectra for instance in disks with an inner dust cavity, which should also affect mostly the shorter MIRI wavelengths. Use of these relations requires a careful inspection of each MIRI spectrum against these effects. These relations should be most reliable where they have been fitted, i.e. at log\lacc$/L_\odot$ values between 0.5 and -2, and extrapolation beyond this range should be taken with caution. While the scatter measured in the regressions is comparable to the average accretion variability recorded in T~Tauri stars (0.3--0.5 dex), indicating its possible origin, line measurements and \lacc estimates should be carefully assessed in individual cases to avoid misleading results.

\begin{deluxetable*}{l c c c c c c c}
\tablecaption{\label{tab: lin_fit_params} Linear regression parameters for the updated correlations shown in Figure \ref{fig: AccRelations_updated}.}
\tablehead{HI line & $\lambda$ [$\mu$m] & Line range [$\mu$m] & $ a (\sigma_a)$ & $ b (\sigma_b)$ & Shift & Pearson $R$ &  Residual RMS}
\startdata
(10-7) & 8.76006 & 8.7508--8.7708   & $ 0.99 ( 0.09 )$ & $ -0.67 ( 0.05 )$ & $ -5.3 $ & $ 0.87 $ & $ 0.33 $ \\
(10-7) corr. & 8.76006 & 8.7508--8.7708   & $ 0.92 ( 0.08 )$ & $ -0.55 ( 0.05 )$ & $ -5.3 $ & $ 0.87 $ & $ 0.34 $ \\
(12-8) & 10.5035 & 10.4943--10.511 & $ 1.03 ( 0.12 )$ & $ -0.30 ( 0.07 )$ & $ -5.3 $ & $ 0.79 $ & $ 0.46 $ \\
(8-7) & 19.0619 & 19.0437--19.0839 & $ 1.46 ( 0.14 )$ & $ -0.22 ( 0.07 )$ & $ -5.3 $ & $ 0.86 $ & $ 0.36 $ \\
\enddata
\tablecomments{Linear relations are in the form $\rm{log}(L_{acc}/L_\odot) = a \times [\rm{log}(L_{HI}/L_\odot) - \rm{Shift}] + b$. The line range in the third column indicates the spectral range where the line flux is measured for each line (also shown in Figure \ref{fig: HI_lines_updated}.)}
\end{deluxetable*}

\section{Additional plots and measurements from this analysis} \label{app: additional_plots}
\begin{figure*}
\centering
\includegraphics[width=\textwidth]{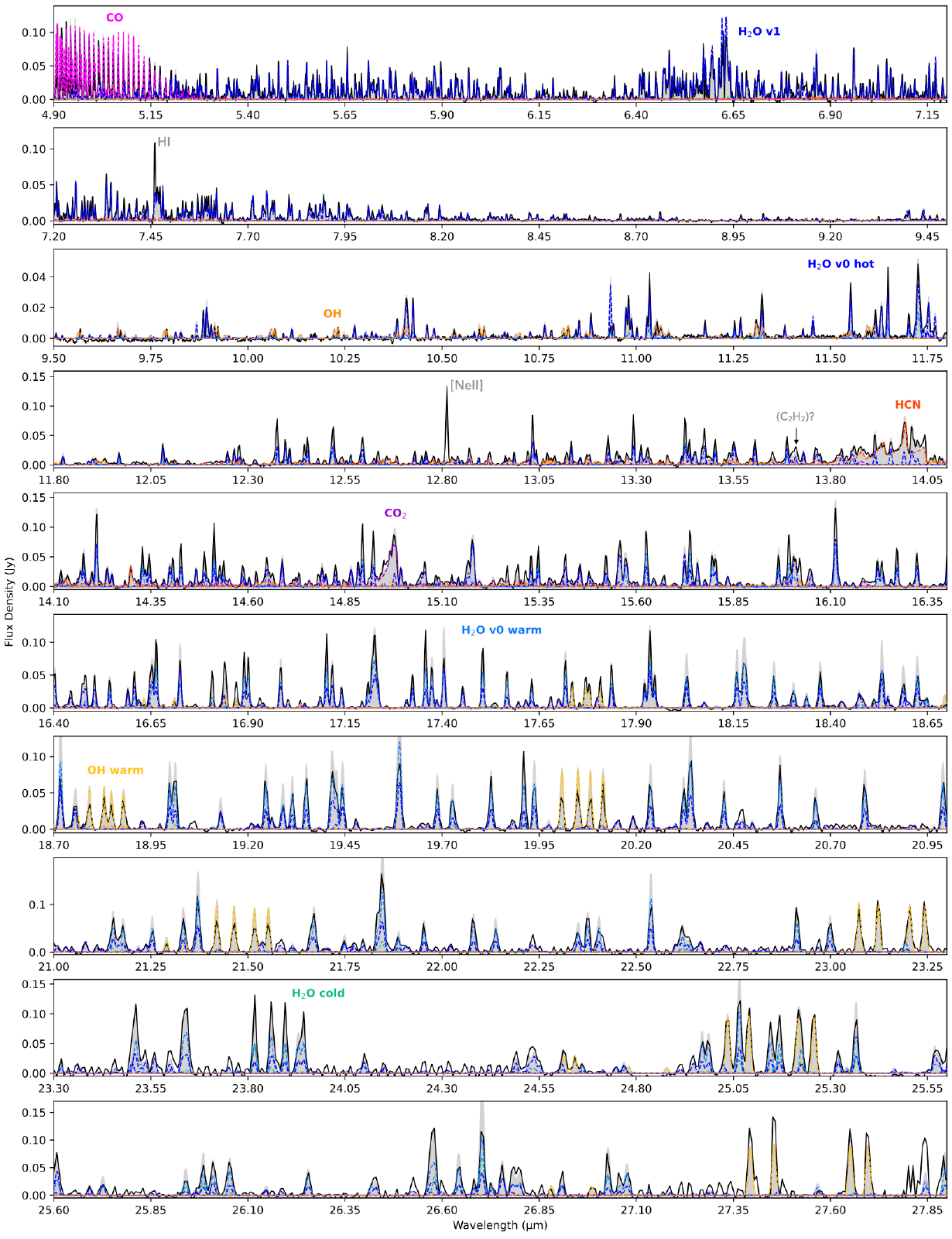}
\caption{Example of full MIRI spectrum obtained in one epoch in this program (September 14), with molecular slab models shown in different colors. Figure made with output models from iSLAT \citep{iSLAT,iSLAT_codev5}.}
\label{fig: full_MIRIspectrum}
\end{figure*}

Figure \ref{fig: full_MIRIspectrum} shows an example of the full spectrum from MIRI for one epoch in this monitoring campaign of DQ~Tau. The figure shows the 3-component best-fit slab models for water emission described in Section \ref{sec: slab_fits} plus representative slab models for other molecules, just for identification of their emission lines and features.

Table \ref{tab: slabfits} reports results from the 3-component best-fit slab models for water emission described in Section \ref{sec: slab_fits} for all the epochs obtained in this monitoring campaign and for the additional four epochs published in \citet{kospal25}, which have been reduced, processed, and fitted in the same exact way as all our MIRI spectra to ensure the most reliable comparison. The slab-fit results from all 13 epochs are shown in Figure \ref{fig: Slab_results_+radii} in the main text. Median uncertainties on water emission best-fit values are approximately 20~K in temperature, 0.1 dex in column density, and 0.1 dex in equivalent emitting mass, similar to previous works \citep[e.g.][]{temmink25,zhang26}. These previous works discussed that the nominal uncertainties may under-estimate the real uncertainty in the multi-component fitting process. The trends found in this work from the multiple epochs of DQ~Tau spectra (Figures \ref{fig: BestTNRM} and \ref{fig: Slab_results_+radii}) support that these uncertainties may generally be reliable, at least when fitting a single disk where the same data quality has been obtained over multiple epochs.

\begin{deluxetable*}{|c|c|c|c|}
\tabletypesize{\footnotesize}
\tablewidth{0pt}
\tablecaption{Multi-component water fit results.}
\label{tab: slabfits}
\tablehead{
Epoch & Hot Component & Warm Component & Cold Component
}
\tablecolumns{4}
\startdata
\begin{tabular}[c]{@{}c@{}}Sep. 11 \\ (B1.1)\end{tabular} &
\begin{tabular}[c]{@{}l@{}}T = 939 \\ N = 1.10$\cdot10^{18}$ \\ A = 0.461 \\ M = 5.73$\cdot10^{-7}$\end{tabular} &
\begin{tabular}[c]{@{}l@{}}T = 559 \\ N = 6.57$\cdot10^{17}$ \\ A = 3.11 \\ M = 2$\cdot10^{-6}$\end{tabular} &
\begin{tabular}[c]{@{}l@{}}T = 229 \\ N = 2.42$\cdot10^{15}$ \\ A = 706 \\ M = 1.92$\cdot10^{-6}$\end{tabular} \\
\hline
\begin{tabular}[c]{@{}c@{}}Sep. 14 \\ (B1.4)\end{tabular} &
\begin{tabular}[c]{@{}l@{}}T = 967 \\ N = 1.05$\cdot10^{18}$ \\ A = 0.515 \\ M = 5.73$\cdot10^{-7}$\end{tabular} &
\begin{tabular}[c]{@{}l@{}}T = 576 \\ N = 6.51$\cdot10^{17}$ \\ A = 3.04 \\ M = 2$\cdot10^{-6}$\end{tabular} &
\begin{tabular}[c]{@{}l@{}}T = 235 \\ N = 2.09$\cdot10^{15}$ \\ A = 706 \\ M = 1.66$\cdot10^{-6}$\end{tabular} \\
\hline
\begin{tabular}[c]{@{}c@{}}Sep. 26 \\ (B2.16)\end{tabular} &
\begin{tabular}[c]{@{}l@{}}T = 889 \\ N = 1.31$\cdot10^{18}$ \\ A = 0.591 \\ M = 8.73$\cdot10^{-7}$\end{tabular} &
\begin{tabular}[c]{@{}l@{}}T = 554 \\ N = 6.22$\cdot10^{17}$ \\ A = 3.29 \\ M = 2$\cdot10^{-6}$\end{tabular} &
\begin{tabular}[c]{@{}l@{}}T = 318 \\ N = 5.89$\cdot10^{14}$ \\ A = 706 \\ M = 4.68$\cdot10^{-7}$\end{tabular} \\
\hline
\begin{tabular}[c]{@{}c@{}}Oct. 1 \\ (B2.21)\end{tabular} &
\begin{tabular}[c]{@{}l@{}}T = 882 \\ N = 1.36$\cdot10^{18}$ \\ A = 0.414 \\ M = 6.38$\cdot10^{-7}$\end{tabular} &
\begin{tabular}[c]{@{}l@{}}T = 574 \\ N = 5.44$\cdot10^{17}$ \\ A = 3.62 \\ M = 2$\cdot10^{-6}$\end{tabular} &
\begin{tabular}[c]{@{}l@{}}T = 279 \\ N = 1.02$\cdot10^{14}$ \\ A = 706 \\ M = 8.11$\cdot10^{-7}$\end{tabular} \\
\hline
\begin{tabular}[c]{@{}c@{}}Oct. 5 \\ (B2.26)\end{tabular} &
\begin{tabular}[c]{@{}l@{}}T = 791 \\ N = 1.94$\cdot10^{18}$ \\ A = 0.384 \\ M = 8.38$\cdot10^{-7}$\end{tabular} &
\begin{tabular}[c]{@{}l@{}}T = 531 \\ N = 7.68$\cdot10^{17}$ \\ A = 3.68 \\ M = 3$\cdot10^{-6}$\end{tabular} &
\begin{tabular}[c]{@{}l@{}}T = 233 \\ N = 2.85$\cdot10^{15}$ \\ A = 706 \\ M = 2.26$\cdot10^{-6}$\end{tabular} \\
\hline
\begin{tabular}[c]{@{}c@{}}Oct. 7 \\ (B2.27)\end{tabular} &
\begin{tabular}[c]{@{}l@{}}T = 861 \\ N = 1.61$\cdot10^{18}$ \\ A = 0.292 \\ M = 5.29$\cdot10^{-7}$\end{tabular} &
\begin{tabular}[c]{@{}l@{}}T = 502 \\ N = 1.00$\cdot10^{18}$ \\ A = 4.04 \\ M = 5$\cdot10^{-6}$\end{tabular} &
\begin{tabular}[c]{@{}l@{}}T = 274 \\ N = 1.37$\cdot10^{15}$ \\ A = 706 \\ M = 1.09$\cdot10^{-6}$\end{tabular} \\
\hline
\begin{tabular}[c]{@{}c@{}}Oct. 11 \\ (B3.32)\end{tabular} &
\begin{tabular}[c]{@{}l@{}}T = 817 \\ N = 1.41$\cdot10^{18}$ \\ A = 0.310 \\ M = 4.92$\cdot10^{-7}$\end{tabular} &
\begin{tabular}[c]{@{}l@{}}T = 471 \\ N = 1.05$\cdot10^{18}$ \\ A = 3.86 \\ M = 5$\cdot10^{-6}$\end{tabular} &
\begin{tabular}[c]{@{}l@{}}T = 281 \\ N = 1.09$\cdot10^{15}$ \\ A = 706 \\ M = 8.71$\cdot10^{-7}$\end{tabular} \\
\hline
\begin{tabular}[c]{@{}c@{}}Oct. 13 \\ (B3.34)\end{tabular} &
\begin{tabular}[c]{@{}l@{}}T = 827 \\ N = 1.85$\cdot10^{18}$ \\ A = 0.200 \\ M = 4.19$\cdot10^{-7}$\end{tabular} &
\begin{tabular}[c]{@{}l@{}}T = 490 \\ N = 1.07$\cdot10^{18}$ \\ A = 3.56 \\ M = 4$\cdot10^{-6}$\end{tabular} &
\begin{tabular}[c]{@{}l@{}}T = 240 \\ N = 2.52$\cdot10^{15}$ \\ A = 706 \\ M = 2.01$\cdot10^{-6}$\end{tabular} \\
\hline
\begin{tabular}[c]{@{}c@{}}Oct. 19 \\ (B3.40)\end{tabular} &
\begin{tabular}[c]{@{}l@{}}T = 833 \\ N = 1.78$\cdot10^{18}$ \\ A = 0.365 \\ M = 7.31$\cdot10^{-7}$\end{tabular} &
\begin{tabular}[c]{@{}l@{}}T = 497 \\ N = 1.11$\cdot10^{18}$ \\ A = 3.94 \\ M = 5$\cdot10^{-6}$\end{tabular} &
\begin{tabular}[c]{@{}l@{}}T = 271 \\ N = 1.40$\cdot10^{15}$ \\ A = 706 \\ M = 1.11$\cdot10^{-6}$\end{tabular} \\
\hline
\begin{tabular}[c]{@{}c@{}}Jan. 29 \\ (K1)\end{tabular} &
\begin{tabular}[c]{@{}l@{}}T = 955 \\ N = 1.03$\cdot10^{18}$ \\ A = 0.661 \\ M = 7.71$\cdot10^{-7}$\end{tabular} &
\begin{tabular}[c]{@{}l@{}}T = 519 \\ N = 1.13$\cdot10^{18}$ \\ A = 3.15 \\ M = 4$\cdot10^{-6}$\end{tabular} &
\begin{tabular}[c]{@{}l@{}}T = 339 \\ N = 2.35$\cdot10^{16}$ \\ A = 27.4 \\ M = 7.26$\cdot10^{-7}$\end{tabular} \\
\hline
\begin{tabular}[c]{@{}c@{}}Feb. 8 \\ (K2)\end{tabular} &
\begin{tabular}[c]{@{}l@{}}T = 903 \\ N = 1.17$\cdot10^{18}$ \\ A = 0.614 \\ M = 8.09$\cdot10^{-7}$\end{tabular} &
\begin{tabular}[c]{@{}l@{}}T = 570 \\ N = 8.08$\cdot10^{17}$ \\ A = 3.47 \\ M = 3$\cdot10^{-6}$\end{tabular} &
\begin{tabular}[c]{@{}l@{}}T = 385 \\ N = 1.04$\cdot10^{16}$ \\ A = 33.3 \\ M = 3.91$\cdot10^{-7}$\end{tabular} \\
\hline
\begin{tabular}[c]{@{}c@{}}Feb. 13 \\ (K3)\end{tabular} &
\begin{tabular}[c]{@{}l@{}}T = 923 \\ N = 1.27$\cdot10^{18}$ \\ A = 0.468 \\ M = 6.68$\cdot10^{-7}$\end{tabular} &
\begin{tabular}[c]{@{}l@{}}T = 545 \\ N = 8.83$\cdot10^{17}$ \\ A = 3.72 \\ M = 4$\cdot10^{-6}$\end{tabular} &
\begin{tabular}[c]{@{}l@{}}T = 356 \\ N = 2.91$\cdot10^{16}$ \\ A = 22.4 \\ M = 7.31$\cdot10^{-7}$\end{tabular} \\
\hline
\begin{tabular}[c]{@{}c@{}}Mar. 2 \\ (K4)\end{tabular} &
\begin{tabular}[c]{@{}l@{}}T = 894 \\ N = 1.36$\cdot10^{18}$ \\ A = 0.492 \\ M = 6.29$\cdot10^{-7}$\end{tabular} &
\begin{tabular}[c]{@{}l@{}}T = 503 \\ N = 1.13$\cdot10^{18}$ \\ A = 3.60 \\ M = 5$\cdot10^{-6}$\end{tabular} &
\begin{tabular}[c]{@{}l@{}}T = 326 \\ N = 1.03$\cdot10^{16}$ \\ A = 55.4 \\ M = 6.45$\cdot10^{-7}$\end{tabular} \\
\enddata
\tablecomments{3-component slab fit parameters as described in Section \ref{sec: slab_fits} and shown in Figure \ref{fig: BestTNRM}. The temperatures are in Kelvin, column densities in cm$^{-2}$, equivalent emitting area in au$^{2}$, and mass in Earth masses.}
\end{deluxetable*}

\begin{figure*}
\centering
\includegraphics[width=\textwidth]{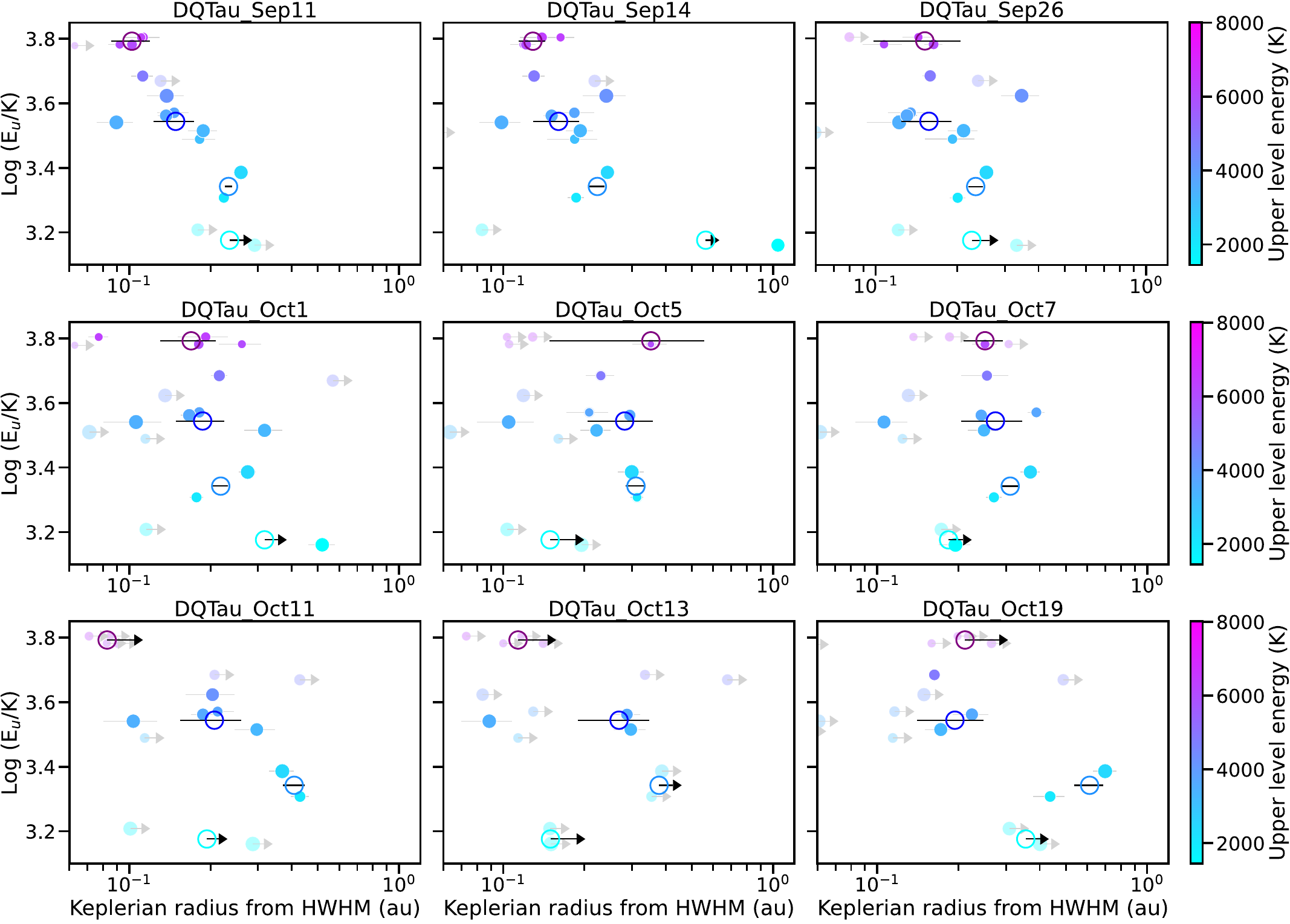}
\caption{ Doppler mapping for water emission lines in all epochs of this monitoring, following Figure 20 in \citet{banz25}.}
\label{fig: Doppler_mapping_all}
\end{figure*}

Figure \ref{fig: Doppler_mapping_all} shows the Doppler maps of Keplerian radii estimated from the line broadening following the analysis presented in \citet{banz25}, similarly to Figure 20 in that work. In this figure, we additionally include the weighted average of Keplerian radii in four bins of upper level energy: the 6000--6500~K lines, the 3000--4000~K lines, the 2000--2500~K lines, and the 1400--1600~K lines, all taken from the list of single lines reported in \citet{banz25} at 14.4--17~$\mu$m. These averages are then used in Figures \ref{fig: Kepl radii} and \ref{fig: Slab_results_+radii} in the main text as $R_{\rm{kepl}}$ at 2200~K, 3500~K, and 6200~K (the approximate center in each bin).

\begin{figure*}
\centering
\includegraphics[width=\textwidth]{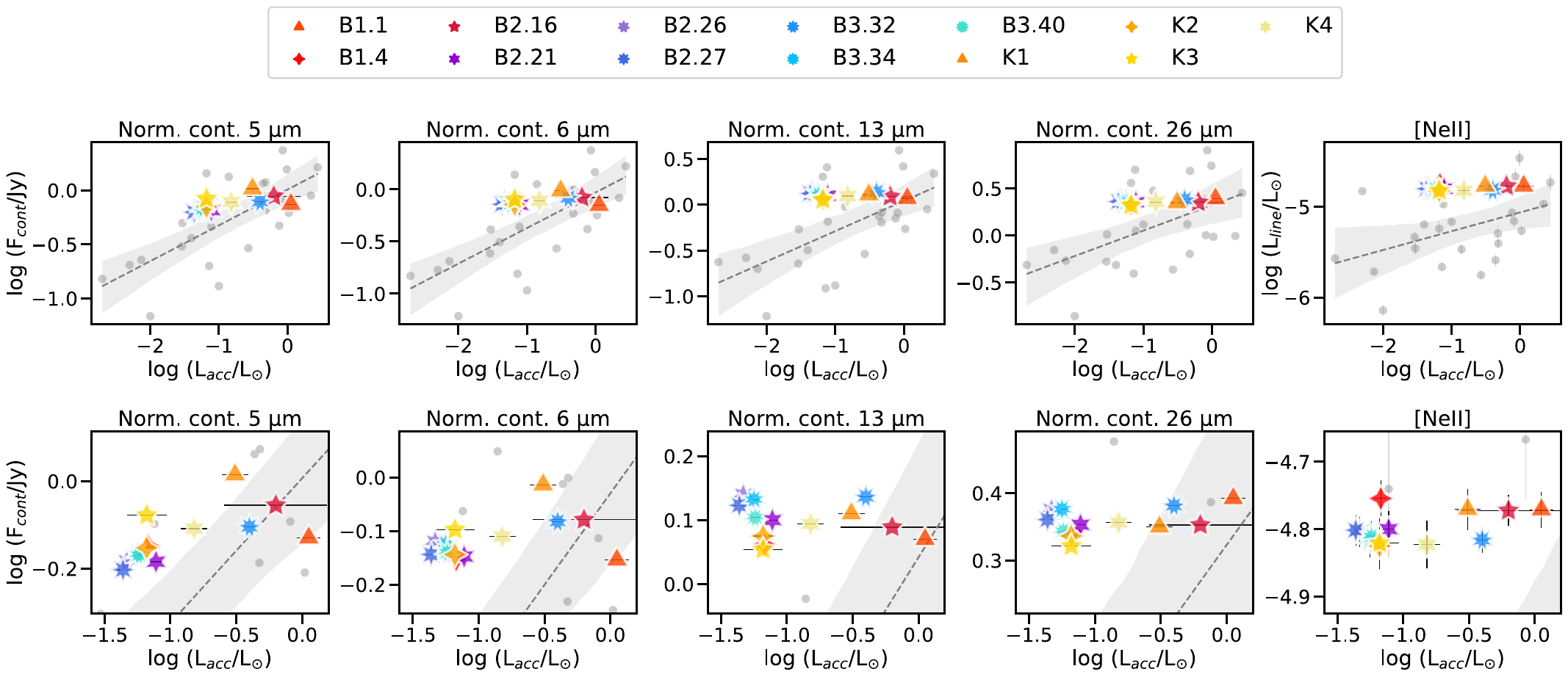}
\caption{Similar to Figure \ref{fig: dqtau_correlations}, here showing the distance-normalized (to 140~pc) continuum flux as a function of accretion luminosity. The epochs from \citet{kospal25} are labeled K1 through 4. The lower rows of plots repeats the upper row but zooming-in on the DQ~Tau values. The [NeII] luminosity is flat with \lacc and generally super-luminous in comparison to the JDISCS sample.}
\label{fig: Lacc_cont_corr}
\end{figure*}

Figure \ref{fig: Lacc_cont_corr} shows accretion trends for the continuum emission (normalized to a common distance of 140 pc in all disks) measured at multiple wavelengths in DQ~Tau and in the reference JDISCS sample, similarly to what is done in Figure \ref{fig: dqtau_correlations} with line luminosities.
In comparison with the JDISCS-C1 sample, as with gas emission lines, DQ~Tau does not follow the global correlations in any of the three continuum ranges, not even at 5 $\mu$m where the DQ~Tau continuum follows the accretion bursts. The likely reason is that the 5~$\mu$m continuum emits from a hotter, closer region near the inner disk edge, which better reflects the change in irradiation from the accretion luminosity. The 13 and 26 $\mu$m continua emit from colder dust at larger disk radii and are less sensitive to accretion. In the low-accretion epochs, the continuum of DQ~Tau tends to be super-luminous in comparison to the trends defined by the JDISCS-C1 sample. 

We also include in Figure \ref{fig: Lacc_cont_corr} the trend for the [NeII] line near 12.81~$\mu$m, which shows a similar flat trend in the DQ~Tau epochs as observed for other colder disk tracers, \ce{CO2}, the lower-energy \ce{H2O} lines, and the continuum flux at 13 and 26~$\mu$m.

\begin{figure*}
\centering
\includegraphics[width=\textwidth]{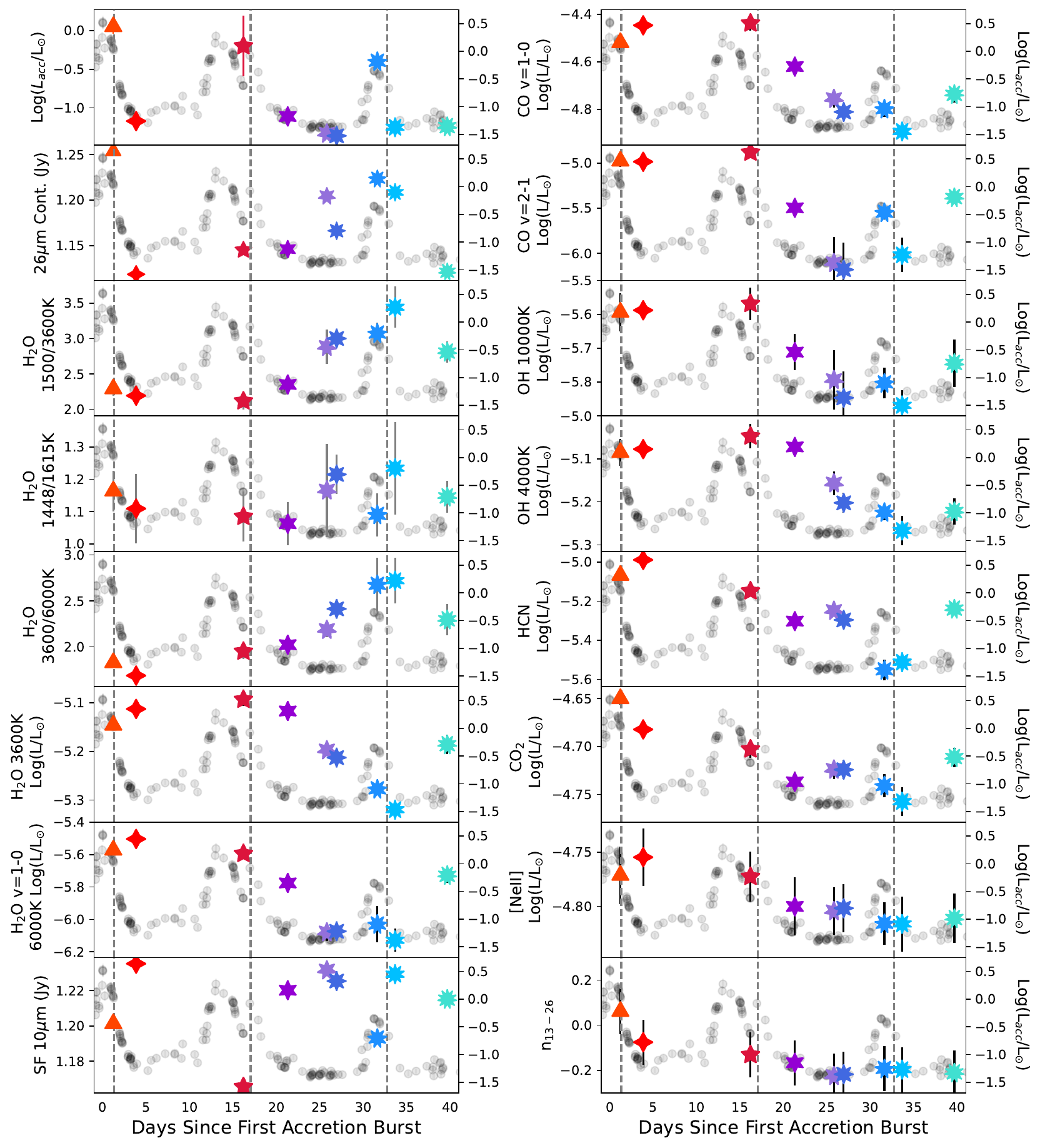}
\caption{Similar to Figure \ref{fig: line_var} but additional line ratios and tracers. The silicate feature strength at 10~$\mu$m is labeled ``SF 10~$\mu$m". }
\label{fig:line_variability}
\end{figure*}

Figure \ref{fig:line_variability} depicts light curves of additional lines to those included in Figure \ref{fig: line_var}. The hotter molecular tracers such as CO, OH, and HCN exhibit similar trends to that of the hot water discussed above in the main text. Emission from [NeII] also follows a similar trend, although with a much lower variation. 

\begin{figure}
\centering
\includegraphics[width=0.45\textwidth]{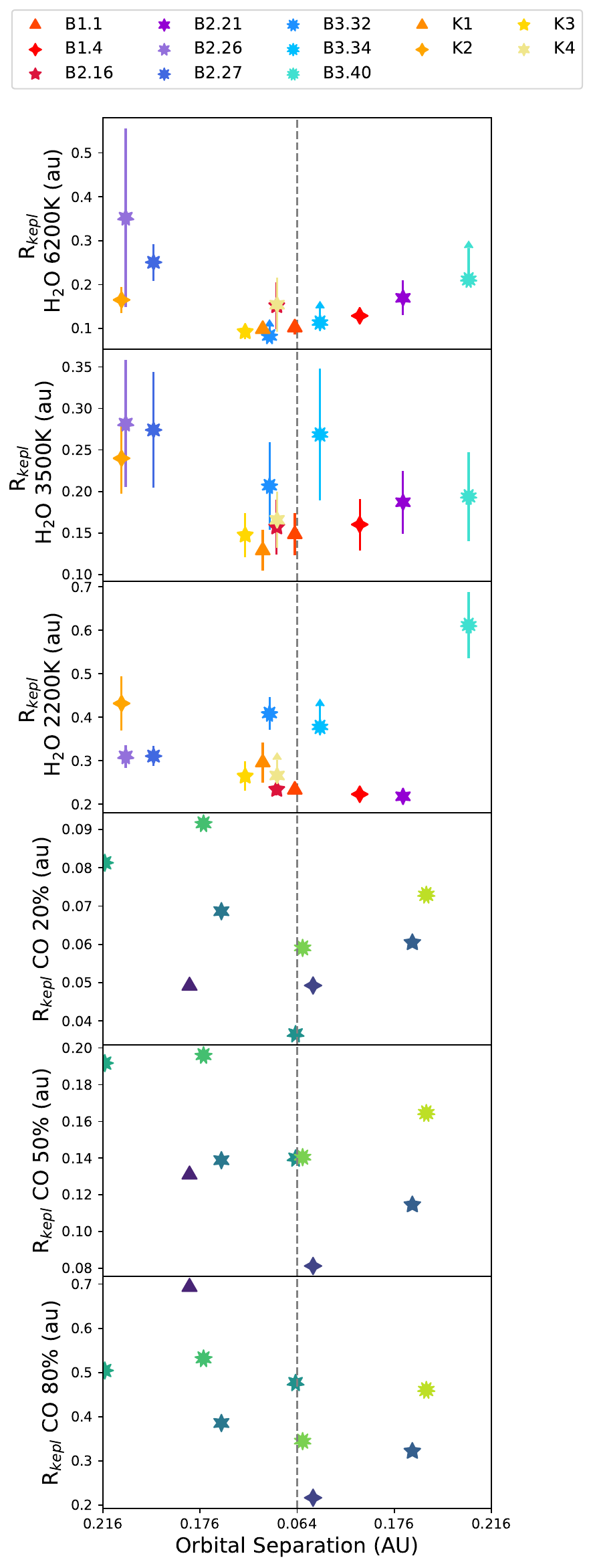}
\caption{Same as Figure \ref{fig: Kepl radii vs time} but as a function of orbital separation from apastron to apastron across periastron (marked with a vertical dashed line).}
\label{fig: dqtau_separation_Vs_radii_zoom}
\end{figure}

\begin{figure}
\centering
\includegraphics[width=0.47\textwidth]{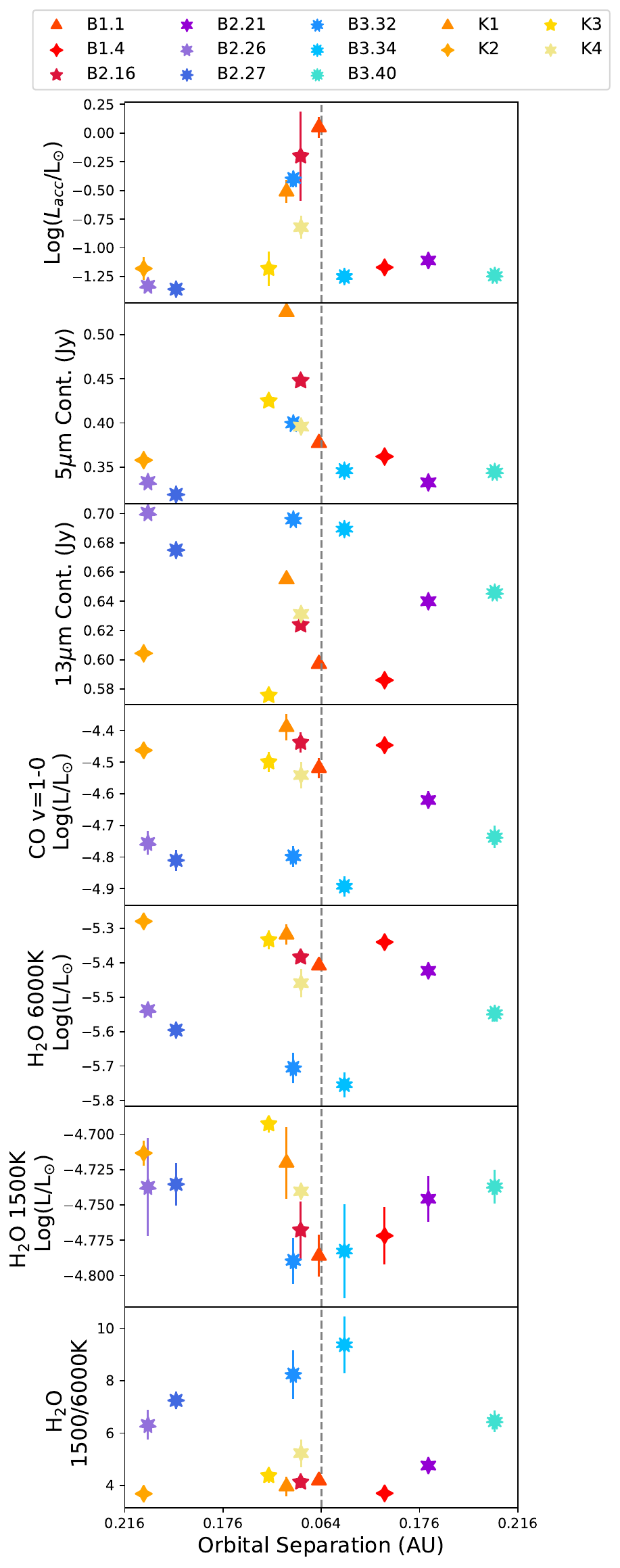}
\caption{Same as Figure \ref{fig: dqtau_separation_Vs_radii_zoom} but for line luminosities and ratios.}
\label{fig: dqtau_flux_vs_separation}
\end{figure}

Figures \ref{fig: dqtau_separation_Vs_radii_zoom} and \ref{fig: dqtau_flux_vs_separation} show the estimated Keplerian radii and molecular emission luminosities as a function of orbital separation, by converting the orbital phases into orbital separation using the parameters from Table \ref{tab: DQTau props}. We show one full orbital period from apastron to apastron across periastron to show the different behavior some tracers have going into vs going out of periastron, and we also include the four additional epochs from \citet{kospal25}. The accretion luminosity and 5~$\mu$m continuum peak right before periastron, as found in previous work (see Appendix \ref{app: accretion_history}). All the molecular tracers studied in this work generally do not show a single trend as a function of orbital separation. The orbital radius and line luminosity of hot tracers tend to be smaller and higher respectively going into periastron, but the two epochs covering the third weak burst in our monitoring show the opposite trend. The CO radii from the 20\% line peak, those tracing the higher-velocity gas, more closely follow the binary orbit by moving in towards periastron and moving out afterwards, as it is visible also in Figure \ref{fig: disc+cartoon} bottom plot. 
The 1500~K line luminosity shows a peculiar trend with an increase right before periastron followed by a minimum at periastron and a rise afterwards. This modulation is remarkable and should be further investigated in future work by obtaining additional epochs to test its stability over time and across multiple orbits.

\bibliography{DQTau_MIRI}{}
\bibliographystyle{aasjournal}

\end{document}